\documentclass[review]{elsarticle} 

\usepackage{natbib}
\usepackage{graphicx}
\usepackage{booktabs}
\usepackage{amsmath}
\usepackage{bm}
\usepackage{color}
\usepackage{colortbl}
\usepackage[version=4]{mhchem}
\usepackage{siunitx}
\usepackage{longtable,tabularx}
\usepackage{makecell}
\usepackage{hyperref} 
\usepackage{mathrsfs} 
\usepackage{comment}
\usepackage{tikz}
\usepackage{nomencl}
\usepackage{multicol}
\usepackage{tcolorbox}
\makenomenclature
\usepackage{rotating}
\usetikzlibrary{calc}
\usetikzlibrary{shapes.arrows, decorations.markings, arrows.meta, calc, positioning}
\usetikzlibrary{arrows.meta, decorations.markings, patterns, calc, shapes.geometric}

\hypersetup{
  colorlinks = true,
  urlcolor  = blue,
  citecolor = blue,
}

\begin{document}

\title{Data-driven methods for flow and transport in porous media: a review}
\author[label1]{Guang Yang}
\author[label1]{Ran Xu}
\author[label1]{Yusong Tian}
\author[label1]{Songyuan Guo}
\author[label1]{Jingyi Wu}
\address[label1]{Institute of Refrigeration and Cryogenics, Shanghai Jiao Tong University, 800 Dongchuan Road, 200240 Shanghai, China}
\cortext[cor]{Corresponding author. Email: x.chu@exeter.ac.uk}

\author[label2,label3]{Xu Chu$^{*}$}
\address[label2]{Department of Engineering, University of Exeter, United Kingdom}
\address[label3]{Cluster of Excellence SimTech, University of Stuttgart, Pfaffenwaldring 5a, 70569 Stuttgart, Germany}


\begin{abstract}

This review examined the current advancements in data-driven methods for analyzing flow and transport in porous media, which has various applications in energy, chemical engineering, environmental science, and beyond. Although there has been progress in recent years, the challenges of current experimental and high-fidelity numerical simulations, such as high computational costs and difficulties in accurately representing complex, heterogeneous structures, can still potentially be addressed by state-of-the-art data-driven methods. We analyzed the synergistic potential of these methods, addressed their limitations, and suggested how they can be effectively integrated to improve both the fidelity and efficiency of current research. A discussion on future research directions in this field was conducted, emphasizing the need for collaborative efforts that combine domain expertise in physics and advanced computationald and data-driven methodologies.

\end{abstract}

\maketitle

\tableofcontents

\section{Introduction}

Flow and transport in porous media is a fundamental subject in various scientific and engineering disciplines, including hydrogeology, petroleum engineering, environmental engineering, and chemical engineering. Porous media, characterized by their internal structure of interconnected void spaces, facilitate the movement of fluids and the transport of solutes, gases, and particulates \citep{blunt2024research,Bottaro.2019,Nepf.2012}. Understanding these processes is crucial for addressing challenges related to groundwater contamination, oil recovery, CO$_2$ sequestration, and soil remediation.
Porous media are typically classified based on their porosity, permeability, and the nature of their solid matrix. The porous structure can range from well-ordered, such as in synthetic materials in industries, to highly heterogeneous, as seen in natural soils and fractured rock formations. The complexity of these structures necessitates a multi-scale approach to study the flow and transport phenomena, encompassing pore-scale interactions to field-scale processes.


The transport of solutes and particles within porous media is influenced by advection, diffusion, and dispersion mechanisms. 
Understanding these mechanisms is essential for predicting the fate and transport of contaminants, optimizing enhanced oil recovery techniques, and designing efficient filtration systems.
Heat transfer in porous media is another critical aspect that intersects with flow and transport processes. The interplay between thermal and mass transport can significantly influence the behavior of multiphase systems, such as in geothermal reservoirs and thermal energy storage systems. 


Advancements in experimental techniques and numerical simulations have greatly enhanced our ability to investigate and predict flow and transport in porous media. X-ray computed tomography (CT) \citep{taghizadeh2023x}, nuclear magnetic resonance (NMR), and microfluidics \citep{Terzis.2019} are some of the tools that provide detailed insights into pore-scale processes. 
The rapid advancement of high-performance computation (HPC) including heterogeneous HPC enables high-fidelity single- \citep{lozano2014effect,lozano2020cause,Pandey2018,Foll2019,Pandey2017,pandey2020,mceligot2018internal,Chu.2016b,Chu.2016c} and multi-phase simulations of different kinds \citep{yi2022numerical,zhang2024investigation,wang2024investigation,xiao2022evaluation}.
Computational approaches, including pore-resolved methods \citep{wood2020modeling,Chu.2020a,yang2020pore,yang2018investigation,wang2021anassess,chu2022investigation,Chu.2021,Chu.2021b,yang2018numerical,yang2019beavers,chu2021turbulence}, pore network modeling \citep{weishaupt2020hybrid}, and volume-averaged approach \citep{wood2020modeling}, enable the simulation of complex interactions within porous structures. The difficulties of current experimental and high-fidelity numerical simulations, such as high computational costs and challenges in accurately representing complex, heterogeneous structures, can potentially be addressed by state-of-the-art data-driven methods.


\begin{figure}[ht!]
\centering
\begin{tikzpicture}[node distance=2cm]
\foreach \i in {1,...,5} {
  \node[fill=yellow!20, draw=black!60, circle] (input\i) at (0,-\i/2) {};
}
\foreach \i in {1,...,10} {
  \node[fill=blue!15, draw=blue!60, circle] (hidden1\i) at (2,-\i/2+1.25) {};
}
\foreach \i in {1,...,10} {
  \node[fill=blue!15, draw=blue!60, circle] (hidden2\i) at (4,-\i/2+1.25) {};
}
\foreach \i in {1} {
  \node[draw=red!60, fill=red!10, circle] (output\i) at (6,-3/2) {};
}
\foreach \i in {1,...,5} {
  \foreach \j in {1,...,10} {
    \draw (input\i) -- (hidden1\j);
  }
}
\foreach \i in {1,...,10} {
  \foreach \j in {1,...,10} {
    \draw (hidden1\i) -- (hidden2\j);
  }
}
\foreach \i in {1,...,10} {
  \foreach \j in {1} {
    \draw (hidden2\i) -- (output\j);
  }
}
\node[above of=input1, node distance=2cm] {Input};

\node[above of=output1, node distance=3cm] {Output};
\node[above of=hidden11, node distance=0.75cm] {Hidden layer};
\node[above of=hidden21, node distance=0.75cm] {Hidden layer};

\end{tikzpicture}
\caption{A three-layer multi-layer-perceptron (MLP) neural network}
\label{MLP}
\end{figure}
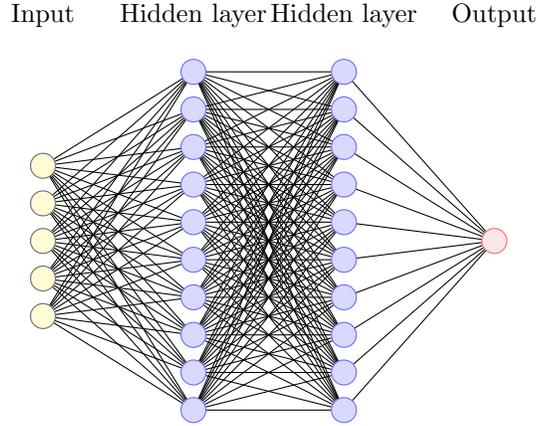

The use of artificial intelligence (AI) has seen a significant increase across various complex applications such as self-driving cars, drug discovery, weather prediction \citep{lam2023learning}, and advanced language processing models like GPT \citep{van2023chatgpt}. Over the past decade, the widespread adoption of ML has been facilitated by the availability of open-source libraries, a robust community, an abundance of data from both physical and numerical experiments, and easy access to HPC. Neural network structures range from simple multi-layer perceptrons (MLP) (Fig.\ref{MLP}) to advanced architectures such as transformers.
Flow modeling has historically relied on first principles, such as conservation laws, which have been the foundation of research for centuries. However, when dealing with multi-scale, multi-phase and multi-physical problems – utilizing scale-resolving simulations based on the Navier-Stokes equations becomes computationally infeasible due to the immense computational demands. As an alternative, approximations of these equations, commonly employed in turbulence modeling, for specific flow configurations are often utilized. Nevertheless, these methods entail substantial costs and time investments, particularly when applied to iterative optimization and real-time control purposes. To address these challenges, substantial efforts have been dedicated to developing accurate and efficient reduced-order models that capture essential flow mechanisms while significantly reducing computational expenses.
In recent years, ML has emerged as a promising approach for achieving dimensionality reduction and developing reduced-order models in fluid mechanics, such as proper orthogonal decomposition (POD), dynamic mode decomposition (DMD), and deep autoencoder (DAE) \cite{chu2024non,chu2023modeling}. ML provides a concise framework that complements and extends existing methodologies. By harnessing data-driven techniques, ML empowers researchers to gain novel insights into complex flow behaviors while maintaining a manageable computational burden. 


The integration of ML into fluid mechanics represents a transformative development, enhancing both modeling and control of fluid flows \citep{duraisamy2019turbulence,wang2023first,Chang.2018,chu2018computationally,beck2019deep,beck2023toward,lozano2023machine}. This includes the deployment of ML across various aspects of fluid dynamics, including enhancements to Reynolds-Averaged Navier-Stokes (RANS) simulations \citep{wu2018physics}, improvements in Large Eddy Simulation (LES) \citep{Beck.2019,beck2021perspective}, innovations in flow control, and the burgeoning field of Physics-Informed Neural Networks (PINN) \citep{raissi2019physics}. These advancements are particularly crucial in applications involving complex porous media, where traditional methods often fall short.
Machine learning augments RANS models, which are pivotal for engineering applications involving turbulent flows. By integrating ML, these models gain enhanced predictive accuracy \citep{ling2016reynolds}. Techniques such as supervised learning can train models on discrepancies between traditional RANS outputs and experimental data, thereby tuning the turbulence models to reflect actual flow conditions more accurately. 
LES is critical for capturing complex fluid behaviors by resolving large-scale flow structures while modeling smaller, more turbulent scales. ML, particularly deep learning models like Convolutional Neural Networks (CNNs), enhances LES by predicting subgrid-scale stresses directly from the resolved scales. This application not only improves the accuracy of LES but also makes it computationally less demanding, which is vital for simulations involving intricate geometries like those found in porous media.

Active flow control techniques, which aim to manipulate the flow to achieve desired outcomes (such as lift increase or drag reduction), are significantly enhanced by ML \citep{vinuesa2024perspectives,vinuesa2022enhancing}. Reinforcement learning, a type of ML, is effectively used to optimize control strategies dynamically. For example, controllers can adjust the operation of actuators in real-time based on feedback from flow sensors, thereby optimizing performance continuously under varying conditions.
ML also revolutionizes passive flow control, which uses fixed devices to alter flow characteristics without mechanical adjustments. Using algorithms like genetic algorithms or deep reinforcement learning, designers can simulate numerous configurations of devices such as vortex generators or dimples to ascertain the most effective setup. This method drastically reduces experimental trial and error by predicting performance impacts through simulation, leading to optimized designs with enhanced aerodynamic properties.




Causal analysis, another critical application, involves understanding the cause-and-effect relationships within fluid flow systems \citep{lozano2020causality,liu2021simulation,Liu2023,liu2023interfacial,liu2023large,liu2024simulation}. Techniques such as Granger causality and transfer entropy \citep{wang2021information,wang2022spatial,chu2022investigation} are used to identify causal links between different flow variables, providing deeper insights into the underlying dynamics of the system. This information is crucial for developing predictive models and designing effective interventions in fluid mechanics applications.

In conclusion, the incorporation of data-driven methods in the study of flow and transport in porous media represents a potential paradigm shift in the field. These approaches offer unprecedented potential for unraveling the complexities of fluid mechanics, enabling more accurate predictions, efficient computations, and innovative solutions to longstanding challenges. By bridging the gap between traditional computational/experimental techniques and modern data science, this review aims to highlight the transformative impact of data-driven methods and inspire further research and application.


\section{Image-based approach} \label{sec.Image}
\subsection{Data augmentation}
 The unprecedented volumes of original data from experiments and field measurements are important for machine-learning to extract information and train models. Generally, the four common porous material imaging techniques belong to time-consuming tasks which included X-ray tomography \citep{metzner2015direct}, magnetic resonance \citep{oswald2015combining}, and neutron and positron imaging X-ray tomography \citep{zahasky2018micro}. Besides the experimental data, the large number of original data can be acquired from the data augmentation to learn robust relationships between the inputs and outputs.
 The data augmentation refers to methods for constructing and introducing unobserved data of latent variables which is applicable to the research of flow and transport in the porous media by data-driven methods \citep{van2001art}. 
 
 In the research of porous media, the data augmentation techniques can generally be categorized as supervised data augmentation and unsupervised data augmentation. Under supervised data augmentation, predefined data manipulation methods are usually used to augment the data based on existing data.
 The basic data manipulation includes image cropping, mixing, and geometric transforms. \citet{alqahtani2020machine} rotated the input images and flipped along horizontal and vertical axes to artificially increase the number of training instances. \citet{varfolomeev2019application} cropped two-dimensional images using a kernel size of $496 \times 496$ pixels and a stride of 248 pixels to carry out the data augmentation. \citet{graczyk2023deep} exploit the property that a horizontal flip of the input image does not change the porosity and diffusion coefficient of the porous media. \citet{liu2019case} extended 2-D data augmentation techniques to a 3-D sub-sampling process by regularly or arbitrarily rotating the image to improve the robustness of the system. \citet{guan2018reconstructing} obtained the dataset from micro-X-ray tomography scales of a Betheimer sandstone. The image data was downsampled to $256^3 $ voxels to yield 36864 training images with extraction every 16 voxels. \citet{rabbani2020deepore} inspired by mixing image method and adopted a weighted interpolation method to binary two textures which is exhibited in Fig. \ref{Chapter3.1,Figure.1}. \citet{ge2024data} made a random crop from a 3-D rock image with a size variation between 50$\%$ to 200$\%$ of the original input size as data augmentation. Then, they resized the image with tri-linear interpolation for up-sampling and porosity averaging for down-sampling. 
 
 \begin{figure}[htbp]
  \centering
  \includegraphics[width=\linewidth]{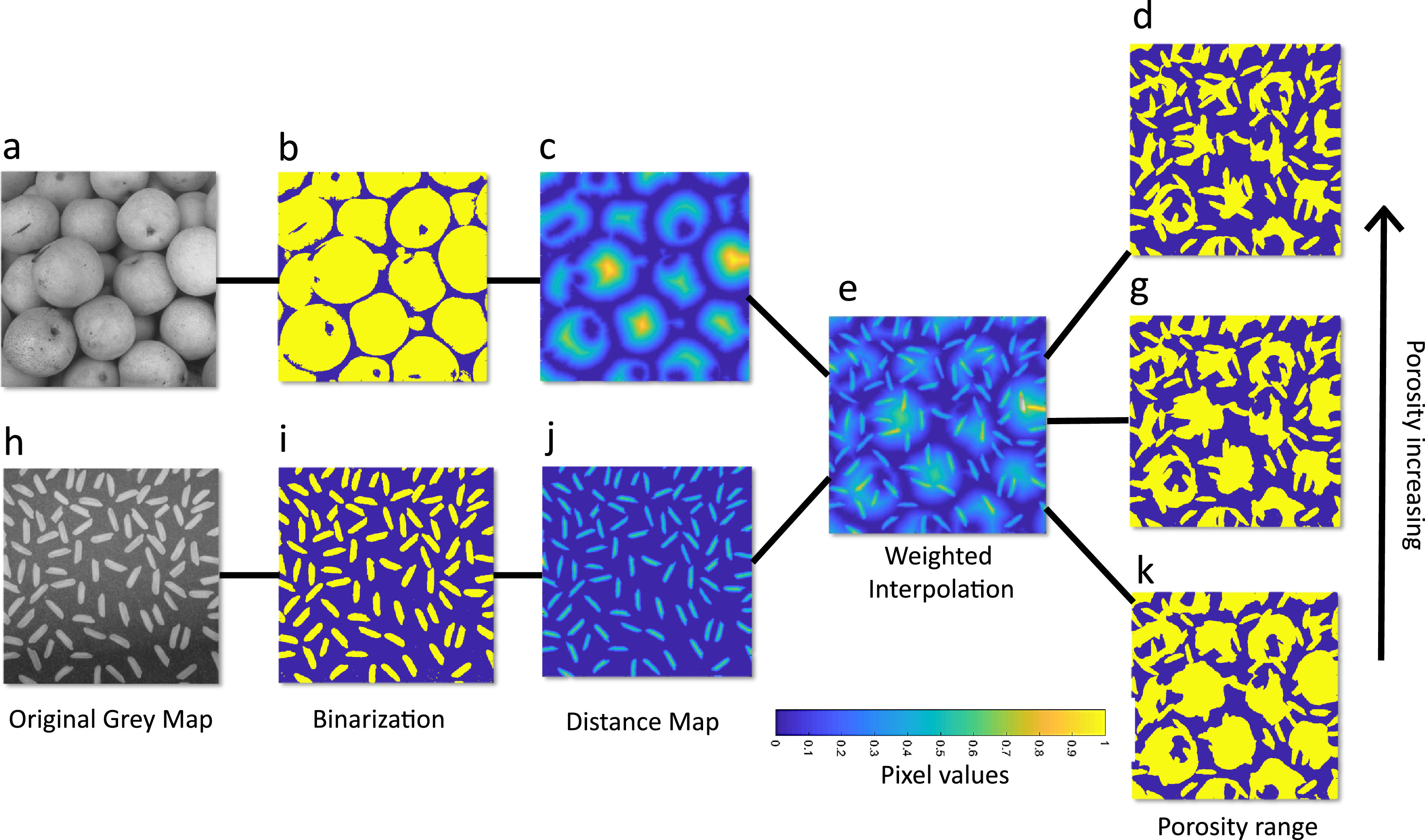}
  \caption{Texture interpolation by weighted averaging of the normalized distance maps, (a and h) original gray maps, (b and i) binarized geometries, (c and j) normalized distance maps of the solid space, (e) equally–weighted average of the distance maps, (d, g, and k) three realizations made by changing the threshold level that controls the porosity \citep{rabbani2020deepore}. Reproduced with permission from Elsevier.}
  \label{Chapter3.1,Figure.1}
  \end{figure}

 The unsupervised data augmentation learns the distribution of the existing data and randomly generates the images that are consistent with the distribution of the training dataset.
 The representative data augmentation method is GAN which is discussed in Chapter \ref{ir} to create artificial image data from the Generator(G) model for data augmentation. 


\subsection{Image segmentation}
After data augmentation, images need to be analyzed by image segmentation.
The image segmentation refers to the division of an image into a number of regions based on features such as colour and greyscale. The porous material images generally contain two or more phases, each of which needs to be detected and separated to analyze and understand the flow and transport characteristics in porous media.
The image segmentation separates the different phases, minerals, or fluids for further detailed investigation about the bulk measurement  \citep{purswani2020evaluation} and the CFD modeling \citep{yang2020pore}. The image segmentation method in the research of porous media can be categorized as image segmentation based on the classic segmentation method and deep learning.

The classic image segmentation method used the predefined thresholding level and interface characteristics to segment original images. 
\citet{armstrong2015effect} accomplished image segmentation using indicator kriging-based thresholding to characterize the pore morphology and analyze residual oil blob mobilization. The voxel values below the lower limit are labeled as one phase and voxel values above the upper limit are labeled as the second phase. \citet{zhang2019challenges} used the greyscale for quantification and determined a certain thresholding level for image segmentation. Then they analyzed the uncertainties and errors by comparison with image segmentation based on grey scale and experimental results. They concluded that the spatial resolution of the image could affect image segmentation results. \citet{schluter2014image} analyzed every histogram-based global thresholding method which is presented in Fig. \ref{Chapter3.2,Figure.1}a. Based on the analysis of the global thresholding method, they also introduced locally adaptive segmentation methods which are used to smooth phase boundaries and avoid noise errors. They concluded that the major cause of poor segmentation results is image blur and proposed locally adaptive segmentation methods that could remove the image noise and image artifacts.
\citet{brown2014challenges} represented the three phases in model porous media by the phases contact lines instead of grey scale values. They used the watershed algorithm which can be relatively effective in avoiding mis-classification errors at phase boundaries due to partial volume effects \citep{armstrong2012microbial} and presented in Fig. \ref{Chapter3.2,Figure.1}b. The classic image segmentation method contains many limitations including low spatial resolution of the raw images \citep{brown2014challenges}, shadows and artifacts at the inter-phase boundary, and the reconstruction noise \citep{han2016deep}. Considering the limitations and difficulties of classic image segmentation and the development of computility and deep learning models, the deep learning has been widely used in image segmentation \citep{badrinarayanan2017segnet}.

Recently, image segmentation based on deep learning has become more popular in porous material studies and it has been used to distinguish different phases. 
\citet{andrew2018quantified} found that the image segmentation method included classic and machine learning have reasonable agreement and similar accuracy in the images with low-level noise. However, by adding Gaussian blur noise to the input images, classic segmentation misclassified the voxels up to 50$\%$.
 \citet{berg2018generation} demonstrated the performance of the classic image segmentation method compared to deep learning through Waikato Environment for Knowledge Analysis (WEKA). The classic segmentation technique is vulnerable to radial artifacts that generates a plausible segmentation.
\citet{marques2019deep} employed two deep learning methods Seg-Net and U-Net for segmenting and classifying thin rock sections into pores and backgrounds. 
\citet{varfolomeev2019application} estimated performance of 2-D SegNet, 2-D U-net, 3-D U-net for processing micro-CT image of rock samples in image segmentation. They also introduced the application of edge-preserving filtering before the application of deep learning for segmentation and found that pre-processing of the grey scale images can affect segmentation results.
\citet{da2020physical} investigated U-Net, Seg-Net, and Res-Net in image segmentation and found that 3-D U-shaped ResNet performs better among the others in mineral segmentation accuracy and simulation permeability.
\citet{mahdaviara2023deep} examined the performance of 2-D and 3-D U-Net deep learning models for multi-phase segmentation of unfiltered X-ray tomograms of gas diffusion layers with different percentages of hydrophobic polytrafluoroethylene and presented in Fig. \ref{Chapter3.2,Figure.1}c. 
\citet{wang2023machine} applied a random forest-based machine learning algorithm and UNet++ deep learning for image segmentation of 3-D scanned shale images from micro-CT.
\citet{siavashi2024segmentation} used autoencoders for semantic image segmentation, which was defined as the categorization of pixels within an image into semantic categories. Four deep autoencoders, SegNet, UNet, ResNet and UResNet were trained on X-ray images of a brine decane two-phase experiment and presented in Fig. \ref{Chapter3.2,Figure.1}d. After comparison, they concluded that the UResNet algorithm, upon adequate training, contained the best proficiency in segmenting multi-phase flow images.

\begin{figure}[!htbp]
  \centering
  \includegraphics[width=\linewidth]
 {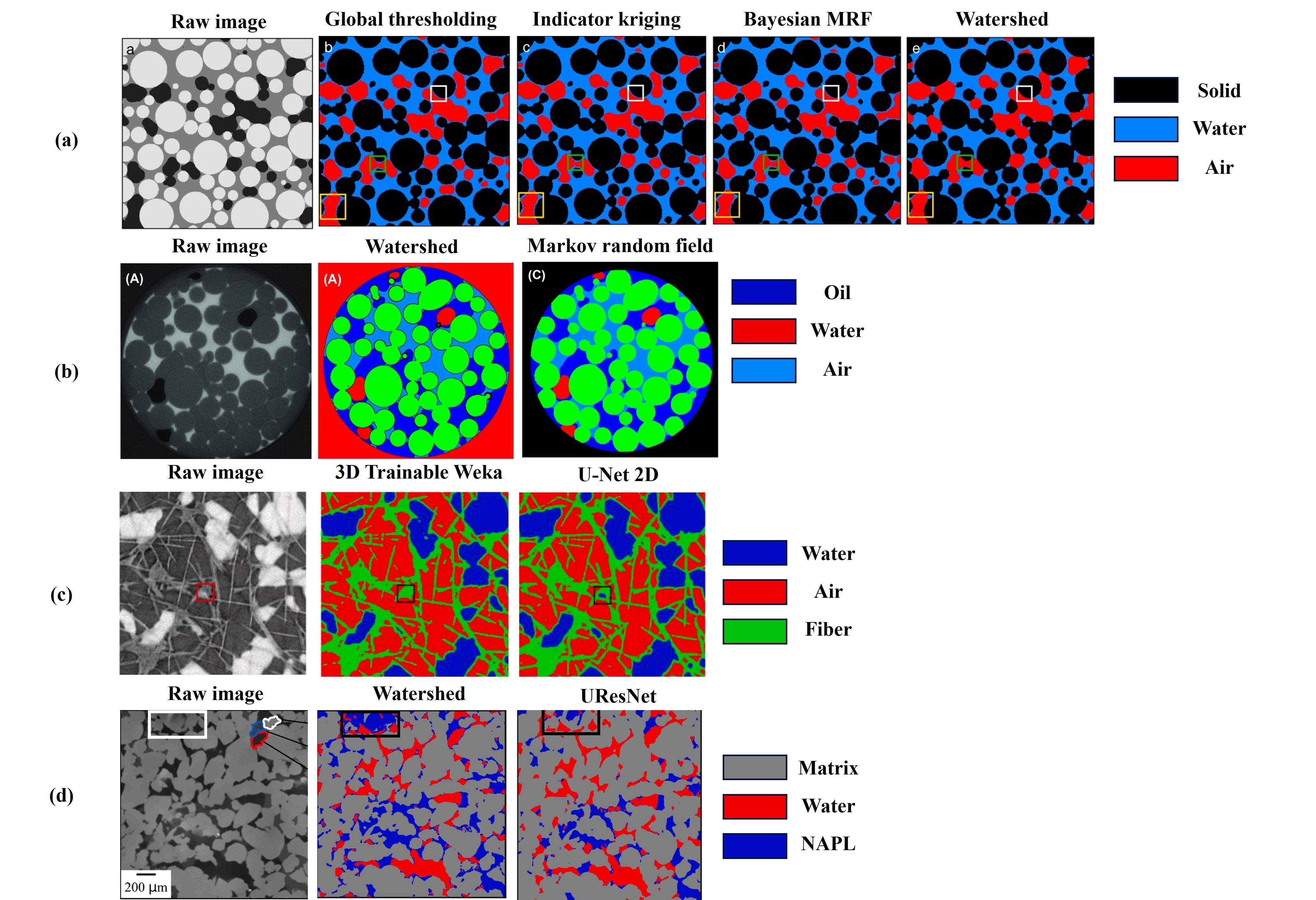}
  \caption{Examples of porous material image segmentation using different methods. (a) Global thresholding and locally adaptive segmentation methods \citep{schluter2014image}. Reproduced with permission from John Wiley and Sons,  (b) Watershed method\citep{brown2014challenges}. Reproduced with permission from John Wiley and Sons. (c) 2-D U-Net and 3-D U-Net \citep{mahdaviara2023deep}. Reproduced with permission from Elsevier. (d) UResNet\citep{siavashi2024segmentation}. Reproduced with permission from Elsevier.}
  \label{Chapter3.2,Figure.1}
  \end{figure}

\subsection{Image reconstruction}
\label{ir}
  To acquire the morphological, topological, and flow properties of porous media, we need not only the division of an image into a number of regions through image segmentation, but also reliable tomography images of porous media. 
  The experimental images are generally 2-D image data which contains low-resolution cells. The image reconstruction technique and the super resolution technique should be introduced. The image reconstruction technique provides a 2-D/3-D complete structure realistic-look image from the experimental image data for predicting effective thermal conductivity, permeability, and interfacial heat transfer coefficient \citep{blunt2017multiphase,bodla20143d}. The super resolution technique is presented in the Chapter \ref{sr}.

  The image can be reconstructed through classic reconstruction methods, which included stochastic reconstruction methods \citep{adler1990flow,quiblier1984new,yeong1998reconstructing}, (non) ballistic procedures \citep{vold1960sediment,visscher1972random,jullien1987simple}, process-based reconstruction methods \citep{bakke19973} and discrete-element methods \citep{cundall1979discrete,van2016machine}. The classic reconstruction methods are mostly time-consuming and in some cases overestimate or underestimate the dynamical properties \citep{andra2013digital}. The deep learning methods show more realistic image reconstruction \citep{baraboshkin2020deep}.
  
  Furthermore, GAN as an unsupervised deep learning methodology, has been designed to perform fast sampling from the learned probability density representation allow full parallel generation, and become the dominant model of image reconstruction, making the generator(G) appropriate to generate large volumetric 3-D images of porous media. 
  \citet{mosser2017reconstruction} developed a Deep Convolutional GAN(DCGAN) to generate the 3-D images which has been presented in Fig. \ref{Chapter3.3,Figure.1}a. 
  The G and D networks are both fully CNNs. G tries to generate 3-D images, while D tries to distinguish between fake and real images' probability density. D tries to label any received sample correctly, meanwhile, G learns how to create images to fool D. In the process of the adversarial battle between G and D, the visual inspection, different morphological, statistical, and dynamical parameters are evaluated the reconstruction performance \citep{goodfellow2014generative,goodfellow2016deep}.
  In some cases, the training dataset of images is incomplete which is difficult to reconstruct. \citet{feng2018accelerating} proposed the conditional GAN(CGAN) to recover the full 2-D images and generate the 3-D image through stacking in Z-direction which has been presented in Fig. \ref{Chapter3.3,Figure.1}b. Three-step sampling(TSS) is a powerful layer-by-layer reconstruction method. After TSS, the sampling images are available and reconstructed using CGAN considering the continuity and variability of adjacent layers in 3-D porous media. The coupled TSS-CGAN method is 750 times faster than the classic multiple-point statistics method(MPS).
  In order to better style and levels of detail in synthetic images, the style-based GAN is proposed which has been presented in Fig. \ref{Chapter3.3,Figure.1}c. D consists of a down-sampling block and fully connected layers which add more feature maps to the output. \citet{hajizadeh2012algorithm} applied image quilting on porous media images created by StyleGAN. \citet{shams2020coupled} developed a coupled GAN and auto-encoder(AE) networks to reconstruct 3-D porous media images which have been presented in Fig. \ref{Chapter3.3,Figure.1}d. The AE consists of an encoder, code, and decoder. The encoder extracts features of input images and compresses them to code. The coupled GAN-AE algorithms provide more realistic results compared with the Modified Joshi-Quiblier-Adler reconstruction(MJQA) \citep{bodla20143d,li2023improved}. In addition, the resolution of generated images of the GAN-AE model is four times the resolution of the original coarse representation. Because of the GAN model's low reconstruction efficiency, \citet{feng2020end} proposed a general end-to-end deep learning-based 3D reconstruction framework called the state-of-the-art GAN-based method (Bicycle GAN) which has been presented in Fig. \ref{Chapter3.3,Figure.1}e. The mapping between a 2-D slice and its 3-D structure is learned by the neural network. This method can achieve 3.6$\times$ $10^4$ speedup factor compared with the classical method.
  \citet{zhang2022fast} proposed a deep learning algorithm based on GAN and CNN named LGCNN. In particular, the size of constructed porous media is far larger than previous GAN reconstruction algorithms as much as 3-4 orders of magnitude.
  \citet{amiri2024true} employed the SliceGAN comprised of a 3-D G with 3-D transpose convolution layers, alongside three 2-D discriminators which has been presented in Fig. \ref{Chapter3.3,Figure.1}f. During training, the G creates 3-D images and the random slices are fed into the D. This model can capture the heterogeneity and variability that cannot be captured by the limited field of SEM tomography and sidestep the 3-D reconstruction constraints. The images produce by Generator has been extensively used to unsupervised data augmentation in \citet{argilaga2023fem,argilaga2023fractal} and \citet{corrales2022wasserstein}.

\begin{figure}[!htbp]
  \centering
  \includegraphics[scale=0.435]
  {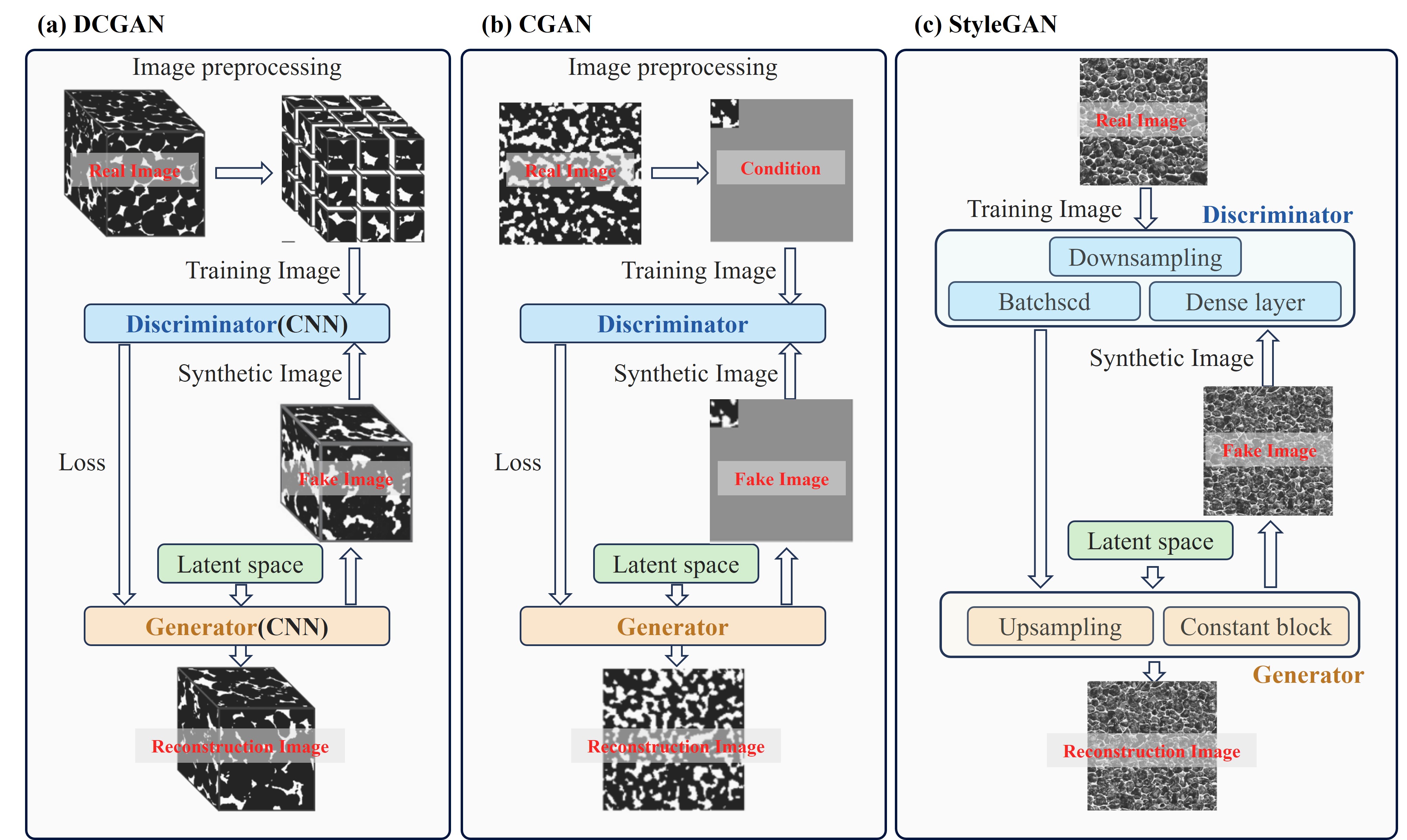}
  \includegraphics[scale=0.45]
  {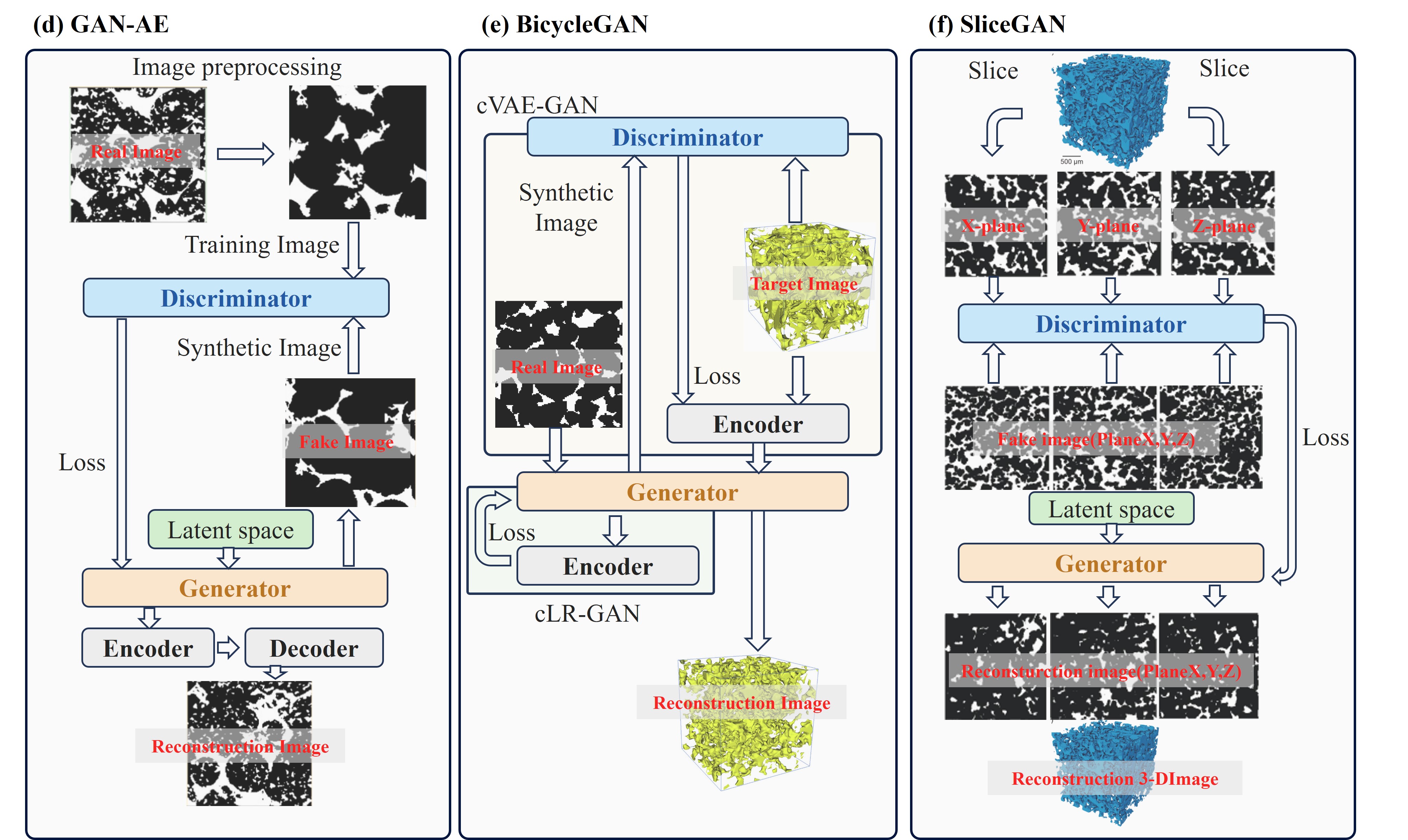}
  \caption{Six GAN models that have been used to reconstruct tomographic images of geo porous material. (a) DCGAN \citep{mosser2017reconstruction}. Reproduced with CC BY 4.0. (b) CGAN \citep{feng2018accelerating}. Reproduced with permission from Elsevier. (c) StyleGAN \citep{hajizadeh2012algorithm}. Reproduced with permission from Springer Nature. (d) GAN-AE \citep{shams2020coupled}. Reproduced with permission from Elsevier. (e) BicycleGAN \citep{feng2020end}. Reproduced with permission from Elsevier. (f) SliceGAN \citep{amiri2024true}. Reproduced with CC BY 4.0.}
  \label{Chapter3.3,Figure.1}
  \end{figure} 

  In addition to the deep learning GAN-based model, many studies have utilized other algorithms commonly used in computer vision. 
  In the research of \citet{zhang20223d}, a recurrent neural network (RNN) based model, namely 3D-PMRNN has been proposed to solve the 2D-TO-3D reconstruction of porous media. This model requires only one 3D sample at least and reconstructs the 3D structure layer-by-layer. The result shows that the synthetic layers of 3D-PMRNN show better diversity than BicycleGAN \citep{feng2020end} and the reconstruction efficiency has been improved compared to the GAN-based model.
  \citet{papakostas2020nature} studied six nature-inspired optimization algorithms and concluded that there are needs for more descriptive and high-order functions for measuring the material's phase distribution to increase the fidelity of the reconstructed microstructure.

  \subsection{Super resolution}
  \label{sr}
  The super resolution technique involves the inference of a high-resolution (HR) image from low-resolution (LR) measurements \citep{brunton2020machine}. The quest for HR images has been one of the major pursuits in both experimental and numerical research. The numerical method to simulate the velocity fields within the porous media is generally DNS method which is highly demanding in terms of computational time and mesh requirements. 
  Furthermore, the super resolution technique can improve the flow prediction performance effectively instead of increasing spatial grid resolution \citep{zhou2022neural}. In the experimental, the porous media images lack sub-micron structure information which is important to predict the fluid transportation properties \citep{blunt2017multiphase,shams2020coupled} because of the micro-CT probing limitations. Because of the micro-CT resolution limitations, the super resolution reconstruction provides robust approaches to improve resolution and remove noise and corruption based on statistical inference. The super resolution reconstruction enables a realistic resemblance of real HR porous media structure images with a desirable field of view \citep{zhang2018super}. Each of the two studies is described as follows respectively.

  In the study of flow fields in the porous media, the DNS can generally be implemented using the lattice Boltzmann method, finite volume method, or finite element method. In order to capture turbulent flow within porous media precisely, the refined mesh is necessary. The ML is capable of learning the complex relationship between experiments and a cost-effective prediction in the simulation of coarse mesh. The CNN and U-net models are commonly used for flow field prediction, but the prediction accuracy in complex porous structures is low especially the pore flow with small dimensions \citep{srisutthiyakorn2016deep,wu2018seeing,wang2021ml}.
  The super resolution techniques have been introduced and developed to reconstruct fine-scale turbulent flow from coarse velocities \citep{liu2020deep, subramaniam2020turbulence}. Although super resolution has been used in many flow field reconstruction research\citep{esmaeilzadeh2020meshfreeflownet, chung2024turbulence, haokai2024refined}, the flow field in the porous media has not been extensively studied due to the complexities involved in the structure.
  \citet{zhou2022neural} developed the U-net, a successive encoder-decoder network that used the geometry and coarse velocities as the model input with the super-resolution technique. The super-resolution can improve the prediction accuracy and regularize the ill-posedness of this learning problem which has been presented in Fig. \ref{Chapter3.4,Figure.1}a.
  \citet{pawar2024geo} employed vanilla SRGAN and Gg-SRGAN to generate HR concentration profiles which was presented in Fig. \ref{Chapter3.4,Figure.1}b. The results show that the difference between the images reconstructed from Gg-SRGAN and the ground truth solutions generated by the finite element method is low. \citet{doloi2023super} proposed using super-resolution techniques to reconstruct fine-scale oil saturation map of an upscaled-reservoir model from a low-resolution simulation.


  \begin{figure}[!htbp]
  \centering
  \includegraphics[width=\linewidth]  {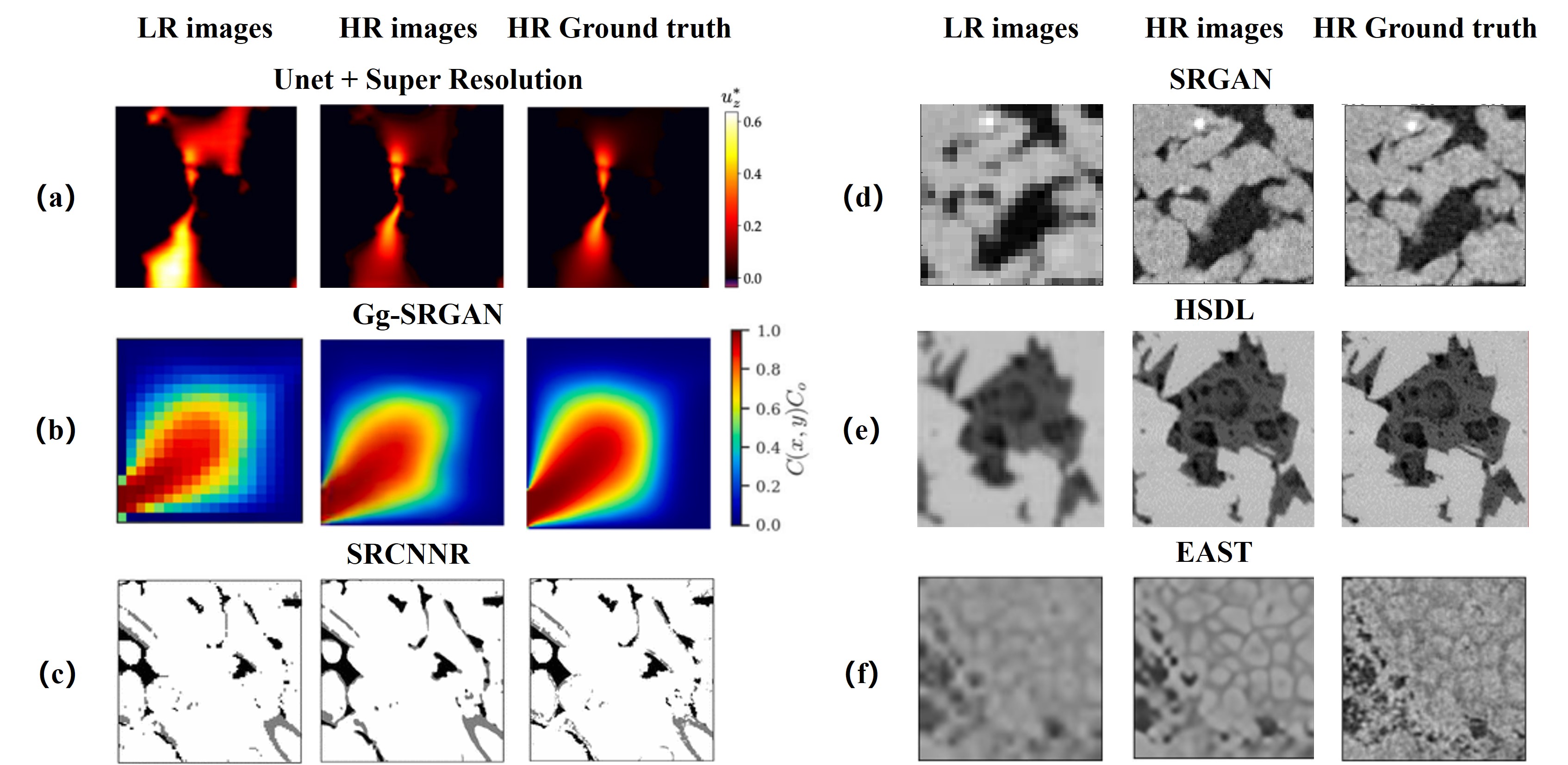}
  \caption{Super resolution examples of porous material reconstruction including the flow fields. (a) The prediction of velocity by UNet+SR \citep{zhou2022neural}. Reproduced with permission from APS. (b) The prediction of concentration by Gg-SRGAN \citep{pawar2024geo}. Reproduced with permission from Elsevier. (c) The porous structure by SRCNNR \citep{wang2020boosting}. Reproduced with permission from John Wiley and Sons. (d) SRGAN\citep{zhang2023super}. Reproduced with permission from Springer Nature. (e) HSDL\citep{kamrava2019enhancing}. Reproduced with permission from Elsevier. (f) EAST\citep{xing2024efficiently}. Reproduced with permission from Elsevier.}
  \label{Chapter3.4,Figure.1}
  \end{figure} 
  
  In the study of porous media structure, super resolution techniques have shown promise as a way to enhance images of real porous media structures. The super resolution can improve the quality of the LR images by reducing the difference error between LR and HR images \citep{karsanina2018enhancing} or finding the mapping relationship between LR and HR images \citep{zhang2018super}.
  \citet{wang2018porous} used the LR and HR 3-D micro-CT images as input. The LR images can be generated from the cubic spline interpolation by HR images. A CNN model was trained to reconstruct the 2-D layers and stacked those 2-D images to build the 3-D HR images. The developed SRCNN can be used to improve the resolution of 3-D porous media images by the corresponding HR 2-D images.
  They found that the SRCNN model would retain the notable noises in the LR images. In other words, the SRCNN reconstructed images have been shown low perceptual quality and even lost some certain texture features loses.
  \citet{wang2020boosting} further introduced the Super Resolution GAN(SRGAN) to improve their previous work(SRCNN) to be usable various images. The DeepRock-SR dataset \citep{da2019super} was used for training the SRGAN network. The SRGAN network can better reconstruct the grains, pores, edges, and HR-SR images due to significant improvement in perceptual and textural features which have been presented in Fig. \ref{Chapter3.4,Figure.1}c. \citet{janssens2020computed} used U-Net generator to reconstruct HR images and compared the porosity and permeability in reconstruction results with experimental results to validate the super resolution model. In particular, two-phase displacement images were used to train U-Net GAN, and the image resolution is improved.
  \citet{zhang2023super} proposed a 3-D reconstruction method SRCSGAN based on ConSinGAN \citep{hinz2021improved} and residual networks for the SR reconstruction of porous media. They also compared the  SNESIM \citep{strebelle2002conditional}, MSPGAN \citep{zhang20223d}, and ConSinGAN and found that SRCSGAN has been proven its accuracy in the SR reconstruction by sandstone and shale samples. Meanwhile, SRCSGAN can save and reuse the training model after the first training which has shown its high efficiency in multiple reconstructions of porous media and presented in Fig. \ref{Chapter3.4,Figure.1}d. \citet{kamrava2019enhancing} generated an augmented dataset of 2-D porous media CT images by cross correlation-based simulation(CCSIM). The hybrid stochastic deep-learning algorithm was proposed. The result confirmed that the HSDL model enabled a better estimation of the physical properties of complex porous media and presented in Fig. \ref{Chapter3.4,Figure.1}e. \citet{zhao2023enhancing} consumed that the result of super resolution was determined by not only voxel size but also the porous media feature such as pore throat. They used the ratio of pore throat size and voxel size to group training set and registered micro-CT as input instead of synthetically down-sampled images. The combination of those two techniques can produce images with better sharpness.
  \citet{xing2024efficiently} introduced self-attention and channel attention mechanisms to undergo structural optimization and proposed a novel Efficient Attention Super Resolution Transformer(EAST) model. The self-supervised fine-tuning approach can enhance the model robustness and resist noise and blur interference and the result is presented in Fig. \ref{Chapter3.4,Figure.1}f.


\section{Data-driven modeling and prediction}

\begin{figure}
    \centering
    \includegraphics[width=0.7\linewidth]{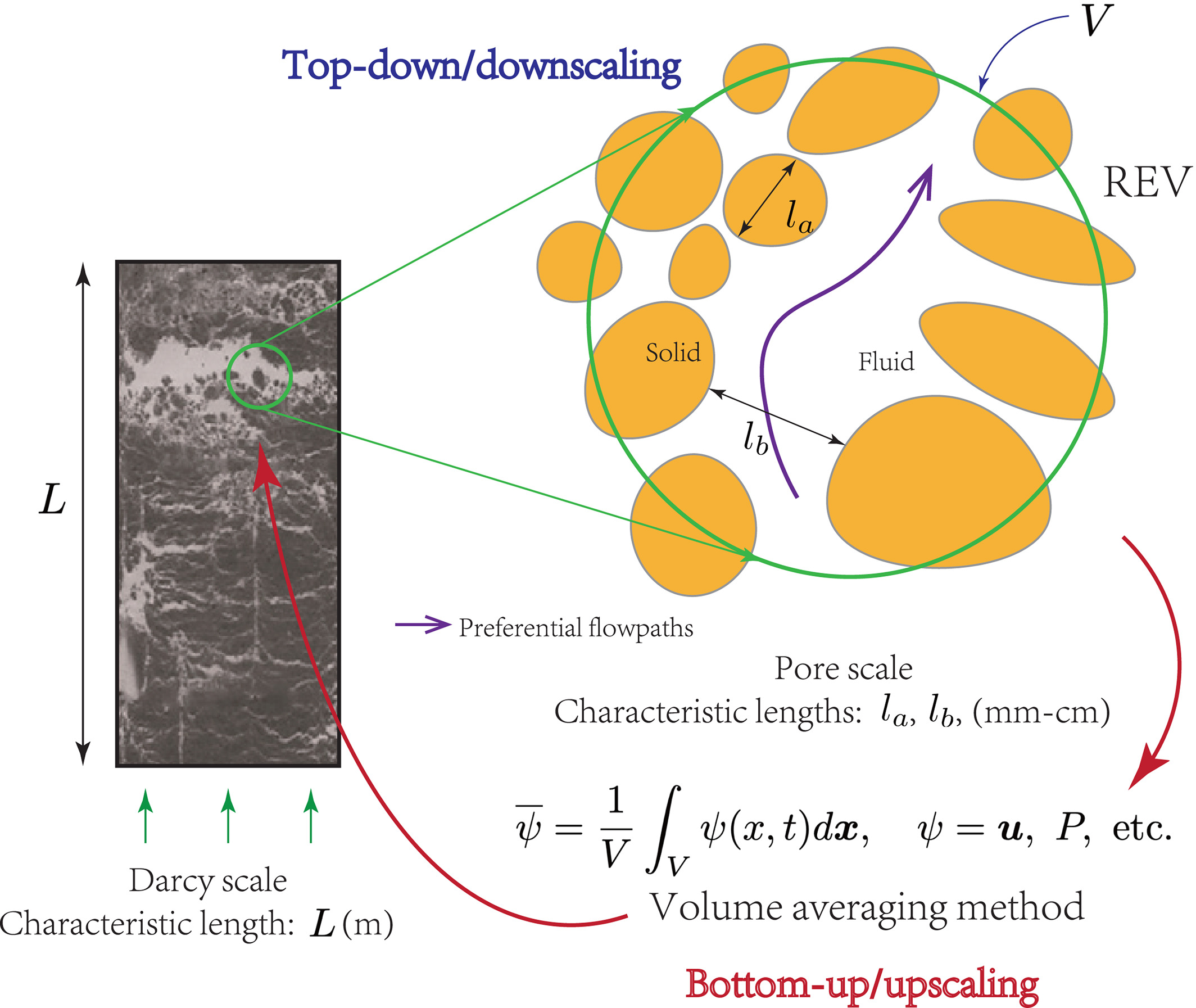}
    \caption{Relationship between macroscopic flow and pore-scale flow \cite{Yang.2021Recent}. Reproduced with permission from John Wiley and Sons.}
    \label{fig:Relation}
\end{figure}

The processes of flow and heat transfer in porous media are generally characterized by multi-scale. The mass and momentum transport phenomena within porous media are highly complex, as the fluid dynamics evolve across scales that differ by several orders of magnitude in both space and time \cite{Song.2016Apparent}. Consequently, the study of flow and heat transfer in porous media has been a challenging problem in the scientific community. In this section, we focus on the prediction of the typical physical fields of flow and heat transfer within porous media based on the data-driven approach. Specifically, the flow dynamics problems are divided into macroscopic and pore-scale views. The relationship between macroscopic flow and pore-scale flow is illustrated in Fig. \ref{fig:Relation}.

\subsection{Macroscopic modeling and prediction}
 Based on the continuum hypothesis, continuum-scale models are frequently used to describe macroscopic modeling and prediction. In these models, a porous medium is typically treated as a homogeneous and isotropic region, neglecting the pore-scale heterogeneity. Macroscopic conservation equations are solved, which are obtained based on volumetric averaging theories for a representative elementary volume (REV) in which phenomenological parameters are introduced to implicitly account for the microstructures of porous media \cite{Bear.1991Modelling}. These phenomenological parameters include porosity, pore size, specific surface area, tortuosity, permeability and relative permeability, effective diffusivity/thermal conductivity, effective reaction rate, etc. For instance, single-phase fluid flow is often described using the Darcy, Brinkman-extended Darcy, or Forchheimer-extended Darcy equations \cite{Darcy.1856Les,Forchheimer.1901Wasserbewegung}, wherein permeability quantifies the flow capacity of porous media. Continuum-scale models are widely adopted for predicting large-scale transport processes of engineering interest. However, since the average length of a computational element in the continuum-scale models is usually much larger than the typical pore size of a porous medium, the microscale heterogeneity of porous media is neglected in the continuum-scale models. In macroscopic modeling, porous media are characterized by structural macroparameters including the porosity and the permeability. 
 Permeability describes the ease with which a fluid can traverse a medium, and it is a critical transport property for 3D structures. This is of significant interest to the research community, particularly because analytical solutions are generally unavailable except for the most simplified, ideal geometries. Continuum-scale models are promising in terms of computational efficiency and user-friendliness. However, these models require additional input parameters from experiments or empirical equations to solve the momentum and energy equations. Moreover, these models do not facilitate flow visualization. Despite numerous efforts to develop relationships that model the permeability based on the structural characteristics of the domain, a universal relationship remains elusive \cite{Ehrenberg.2005Sandstone,Worthington.1993The}. 
 

Deep learning based models have demonstrated excellent accuracy in predicting flow dynamics and superior computational efficiency compared to traditional continuum-scale models \cite{Zhu.2018}. Through extensive data learning, flow visualization and prediction of heterogeneous macro flow variables are realized. \citet{Tang.2020,Tang.2021} combined a residual U-Net with convLSTM networks to predict the temporal evolution of the saturation and pressure fields in 2-D and 3-D oil production simulations. Their recurrent R-U-Net model was later applied to CO$_2$ storage with the coupled flow and geomechanics \cite{Tang.2022Direct}. 
\citet{Wen.2021Towards} developed an R-U-Net-based surrogate model for CO$_2$ plume migration, where injection durations, injection rates, and injection locations are encoded as channels of input images. Wen et al. \cite{wen2022u,Wen.2023Real-time} combined U-Net and Fourier neural operator by adding convolutional information in the Fourier layer, which yields significantly improved cost-accuracy trade-off. Data-efficient multiphase flow predictions for gas saturation and pressure were built. Deep-learning-based surrogate models are proposed by \citet{ju2023learning} to address the challenge of accurately capturing the impact of faults on CO$_2$ plume migration in subsurface flow problems. Their approach, integrating Graph Convolutional Long Short-Term Memory (GConvLSTM) and MeshGraphNet (MGN), demonstrates the potential of GNN-based methods in accurately predicting gas saturation and pore pressure evolution in reservoirs with impermeable faults, highlighting its effectiveness in modeling complex subsurface flow for CO$_2$ geological storage. \citet{Mo.2019} developed a DL surrogate model that integrates an autoregressive model with a convolutional-NNs-based (CNNs) encoder-decoder network to forecast CO$_2$ plume migration in random 2-D permeability fields.

\subsection{Pore-scale prediction}
\begin{figure}
  \centering
  \includegraphics[width=1\linewidth]{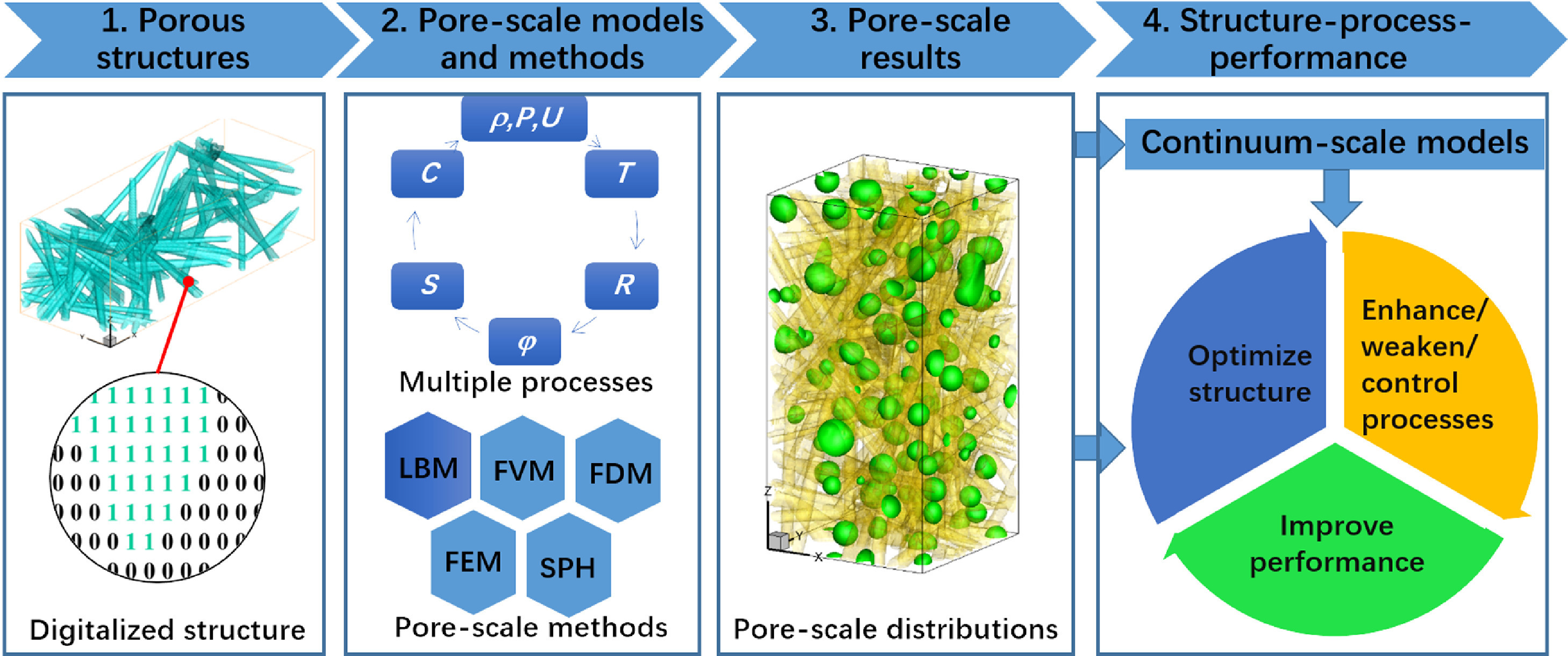}
  \caption{Implementation processes and roles of pore-scale modeling \cite{Chen.2022Pore}. Reproduced with permission from Elsevier.}
  \label{fig: Process}
\end{figure}

\begin{figure}
  \centering
  \includegraphics[width=1\linewidth]{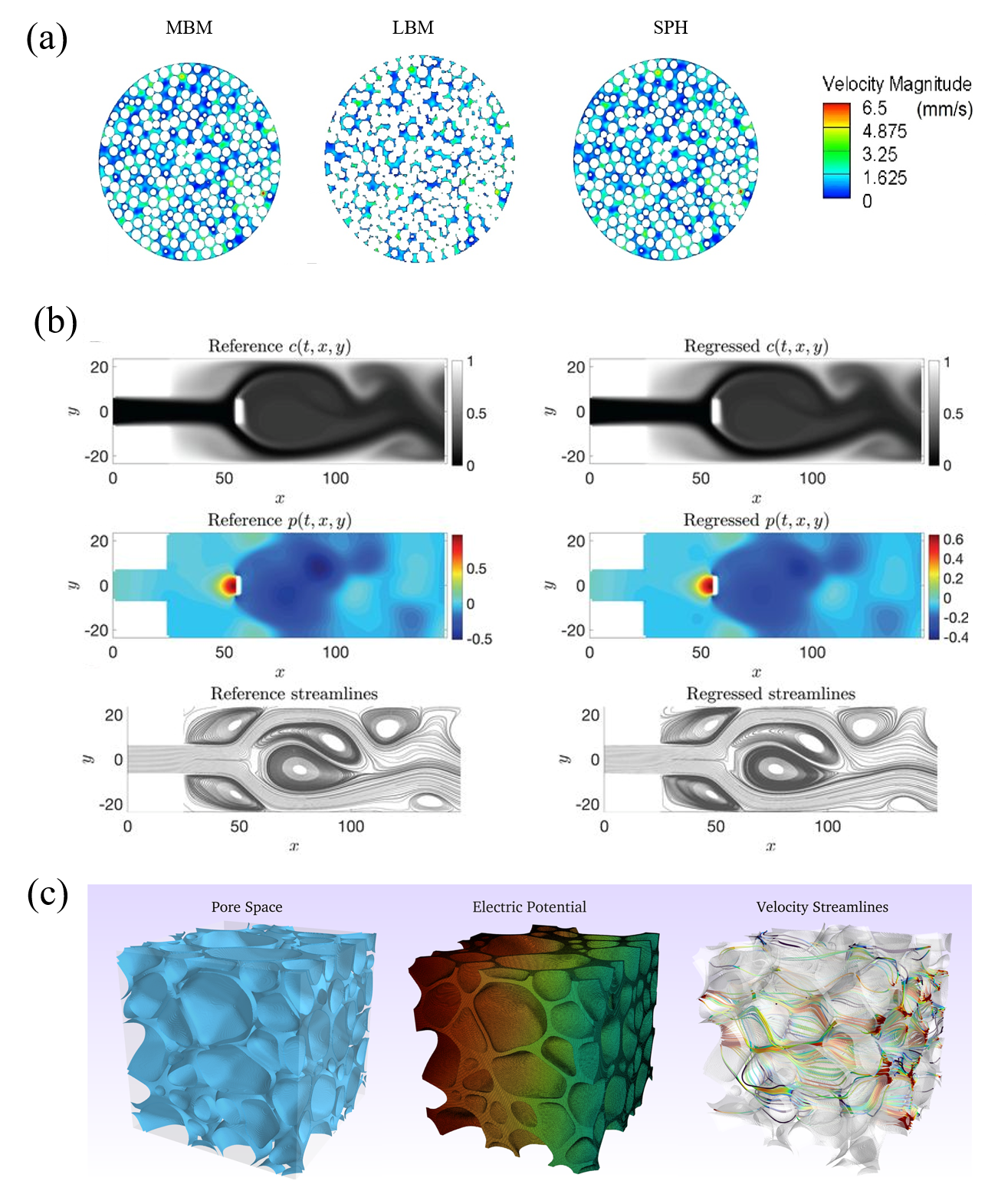}
  \caption{(a) The velocity field predicted by traditional method \cite{Yang.2016Intercomparison}. Reproduced with permission from Elsevier. (b) Hidden states of the system—pressure and velocity fields—obtained by using the ML method based on the data on the concentration field \cite{Raissi.2020Hidden}. Reproduced with permission from AAAS. (c) The electric potential and the streamline plot from the pore-scale prediction by ML method \cite{santos2022dataset}. Reproduced with CC BY 4.0.}
  \label{fig:Pore-scale comparation} 
\end{figure}

Pore-scale models have been extensively used in the porous media study in order to obtain the microscopic flow topologies \cite{Yang.2022Pore,Chen.2013Pore}. 
The pore-scale models provide distribution details of interested variables at the pore scale, enable a direct link of the complex transport processes to the realistic porous structures, and thus can provide a deep understanding of the relationships between structures, processes, and performance. The role of the pore-scale modeling as well as its typical sub-processes is shown in Fig. \ref{fig: Process}.
Various computational methods have been developed and applied to the pore scale to study flow and transport phenomena \citep{Valvatne.2005Predictive}. Considering the complexity of the pore geometries, the first step prior to modeling is to characterize the pore space and structures. Imaging techniques such as X-ray microtomography (XMT) \cite{Wildenschild.2013X-ray} and magnetic resonance velocimetry (MRV) measurements \cite{Seymour.1997Generalized} as well as the image-based methods in Section \ref{sec.Image} have made it possible to obtain accurate 3-D characterizations of the pore geometry at high resolution. Pore-scale modeling can then proceed either by simulating directly on the complex pore geometry, or on a conceptualized pore network that maintains the same topological structure. The first class of model is typically referred to as direct numerical simulations. These approaches include mesh-based method (MBM) \cite{Chu.2018a,Chu.2019}, lattice Boltzmann method (LBM) \citep{Kuwata.2016,Suga.2020,Jin.2015}
, and smoothed particle hydrodynamics (SPH) \citep{osorno2021cross}. The second class of model represents the pore space as a network connected by geometrically simplified pore bodies and pore throats and most commonly takes the form of pore-network models (PNM) \cite{koch2021dumux}. Both flow and transport processes can be represented using either of these approaches. \citet{Chen.2022Pore-scale} compared the velocity field of different pore-scale flow and solute transport simulation methods as shown in Fig. \ref{fig:Pore-scale comparation}a. These approaches can be combined with machine learning to complement each other.  


Using machine learning (ML) models as surrogate models to estimate properties from data is a promising avenue. Current work in pore-scale ML models has shown promising results \cite{Lubbers.2020Modeling,Wang.2021ML-LBM}. \citet{Raissi.2020Hidden} developed hidden fluid mechanics (HFM), a physics-informed deep-learning framework capable of encoding the Navier-Stokes equations into the neural networks while being agnostic to the geometry or the initial and boundary conditions.
Hidden states of the system —pressure and velocity fields—obtained by using HFM method based on the data on the concentration field, as shown in Fig. \ref{fig:Pore-scale comparation}b. The high accuracy that ML models are beginning to demonstrate may allow them to be orders-of-magnitude faster surrogates for simulation flow of effective properties of pore-scale materials, or for initializing simulations into a near-converged state. \citet{Lubbers.2020Modeling} demonstrate their ML-based scale-bridging framework to capture adsorption under nanoconfinement where there is no clear separation of scales. They incorporate atomistic adsorption effects that occur within a nanoconfined pore as simulated accurately by MD into a continuum LBM that is capable of simulating larger scales. Nevertheless, the pore-scale ML subspace remains vastly unexplored, in part due to the complexity of obtaining and processing data. As examples, current works are often limited to models trained with small 3D samples \cite{Wang.2021ML-LBM} or models trained with bigger samples but with restricted geometry type (spherepacks) \cite{santos2021computationally}. Notably, \citet{Prifling.2021Large-Scale} used 90,000 synthetic microstructures from 9 families of distributions downsampled to 963 to perform 3D machine learning. For porous media, datasets of natural examples are beginning to be assembled, but of limited scale and lack flow information \cite{Ditscherlein.2022PARROT}. Large, labeled datasets are usually limited to synthetic examples \cite{Prifling.2021Large-Scale} and/or 2-D flow. A large-scale collection of pore-scale data is critical for developing advanced ML algorithms that generalize to unseen samples from many real-life sources. In order to push the state of the art for ML forward and provide comparisons between ML techniques, it would be very useful to have a large number of labeled data (the results of expensive full-physics simulations) to build models that work in the complexity of real-world pore structures. \citet{santos2022dataset} developed a diverse dataset that represents challenging, complex porous media in the context of 3D simulation, empirical functional forms, and machine learning, as shown in Fig. \ref{fig:Pore-scale comparation}c. This diversity covers dimensions such as porous media lithology, boundary conditions, geometric resolution, and physical processes simulated. They introduced the large-scale collection of images, geometric data, and flow and electrical simulations. 

\subsection{Heat transfer}
\begin{figure}
    \centering
    \includegraphics[width=1\linewidth]{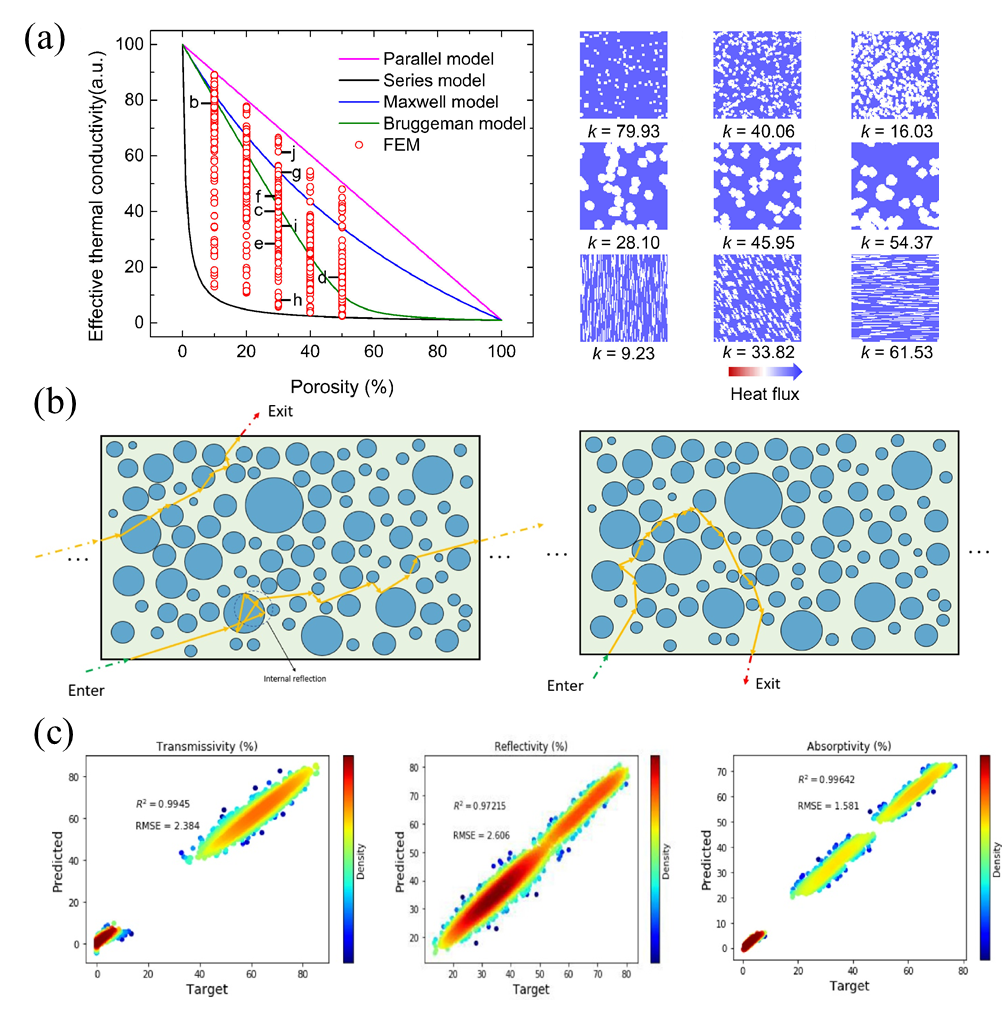}
    \caption{(a) Thermal conductivities and typical porous structures \cite{Wei.2020Machine}. Reproduced with permission from Elsevier. (b) Ray tracing for ray with an entrance angle of 30/60 deg in a porous domain \cite{Hajimirza.2021Learning}. Reproduced with permission from Elsevier. (c) Neural network results with the prediction performance measures, regression plots for transmissivity, reflectivity, absorptivity \cite{Hajimirza.2021Learning}. Reproduced with permission from Elsevier.}
    \label{fig:heat transfer}
\end{figure}

Data-driven machine learning methods have proven effective in studying heat conduction in porous media \cite{RONG2019107861}. \citet{WOS:000445984500080} employed a convolutional neural network (CNN) to predict the effective thermal conductivity of porous media, achieving high accuracy and negligible time required for predicting. By constructing a convolutional encoder-decoder neural network, \citet{WOS:000799016200008} also demonstrated that the temperature field in porous media can be accurately predicted through data-driven analysis. As an image-to-image regression model widely used for image denoising and reconstruction \cite{WOS:000401906900033}, this method yields speedups of several orders of magnitude compared to numerical calculations when predicting the temperature field of the input porous media images. However, a major deficiency of data-driven machine learning methods is that sufficient labeled data is required to constrain the training process and achieve high prediction accuracy \cite{WOS:000799016200008}. Although the labeled training data can theoretically be obtained using any method with good accuracy, in practice, they are commonly obtained from numerical calculations that require solving physics-based partial differential equations. Acquiring substantial amounts of labeled training data is usually challenging, which hinders the application of data-driven machine learning methods for efficient prediction of heat conduction in different porous media \cite{WOS:000453776000028}. 
\citet{xu2023physics} applied physics-informed neural networks to investigate heat conduction in porous media. 
The obtained effective thermal conductivity values for an ensemble of porous media samples had an average relative error of 2.49\%. They also illustrated that physics-informed neural networks could be easily extended to predict nonlinear heat conduction in porous media. \citet{Hajimirza.2021Learning} investigated structural features that have significant effects on thermal transport in porous media and identified five physics-based descriptors to characterize the structural features: shape factor, bottleneck, channel factor, perpendicular nonuniformity, and dominant paths. The proposed descriptors were incorporated into machine learning models to predict the effective thermal conductivity of porous media, and the results were shown to be fairly accurate, as shown in Fig. \ref{fig:heat transfer}a.

 The presence of various-sized pores and particles in porous media results in a complex mechanism of radiative transfer, as rays continually interact with particle-void domains. To accurately demonstrate the radiative properties of structures, a large number of light bundles must be traced to completion or reach the hypothesized conditions. Due to the presence of numerous particles in porous media, the interaction between light, particles, and voids can lead to a variety of patterns and trajectories. This necessitates running many simulations at a high computational cost. ML algorithms have proven to be an efficient alternative for addressing this challenge by learning the nonlinear and complex relationship between the inputs of the problem and the outputs, i.e., the radiative properties. ML can be employed for fast and reliable prediction of radiative properties through appropriate abstraction of the geometric configuration and involved physics, in terms of suitable geometrical and physical features. In recent studies, \citet{Petrasch.2008Tomography-Based} utilized the two-point correlation function and minimum size of representative elementary volume to conduct tomography-based analysis of reticulated porous ceramics. Additionally, other studies have utilized parameters such as tortuosity, chord length distribution, and other geometrical metrics to determine the structure-related transport measures of a 3-D microstructure of porous fuel cell \cite{Cecen.2012Three-dimensional}. In another study \cite{Kang.2019Data}, an artificial neural network (ANN) framework was designed that considered bed height, particle diameter, and emissivity of monodisperse spherical packed beds as input and achieved close approximations for radiative properties compared to Monte Carlo ray tracing (MCRT) ground truth. By utilizing the geometric abstraction, eight geometric features were employed in an ANN algorithm to estimate the optical properties of monodisperse and polydisperse circular and square-shaped pores in porous media. In a more comprehensive study, \citet{Hajimirza.2021Learning} proposed a range of geometrical features for the prediction of the radiative properties of heterogeneous non-overlapping and overlapping circular particles in porous media (Fig. \ref{fig:heat transfer} b and c). A sensitivity analysis was conducted to determine the significance of the features and concluded that the addition of directional features could improve the accuracy of the ANN model prediction. \citet{EGHTESAD2023124742} achieved an enhanced estimation of temperature-dependent radiative properties of heterogeneous porous media by using the Monte Carlo ray tracing algorithm for ground truth data generation. This process involves developing expertly-designed engineered features for geometry abstraction and implementing artificial neural networks for quick and efficient predictions.
 
In addition, due to the complexity of heat transfer mechanisms involving porous media, machine learning methods are also commonly used for model fitting, including heat transfer coefficients for single-phase and two-phase processes. For example, \citet{jahanbakhsh2024physics} developed a generalized predictive model for analyzing heat flux of thin film evaporation within hierarchical structures.













\section{Physics-informed machine learning}
Relying solely on massive amounts of data as direct inputs and simply increasing the depth and width of neural network models to construct complex mapping relationships is not an optimal modeling strategy. Furthermore, data-driven approaches usually require a large amount of high-confidence sample data. However, such methods become impractical when sample data is scarce or when flow field sample data cannot be obtained. Due to the complexity of problems, system characteristics often need to be characterized by partial differential equations (PDEs). Therefore, based on a thorough understanding of flow physics and model architecture, feature extraction of input data can effectively balance model accuracy and generalization capability by reducing input information while ensuring accuracy, simplifying the dimensions and complexity of network models. This approach enhances model interpretability and aids in understanding physical systems. In this section, we will focus on the application of machine learning methods integrated with physical systems in the context of porous media flow and heat transfer.

\subsection{Physics-informed neural network (PINN)}
 The Physical Information Neural Network (PINN) constitutes a novel method within the domain of machine learning, particularly notable in fluid mechanics, as it directly engages with PDEs through neural networks. Originally introduced by Raissi and Karniadakis et al. \cite{raissi2019physics,raissi.2020}, PINN has progressively emerged as a new paradigm for solving both forward and inverse problems in partial differential equations in recent years. Its escalating prominence stems from its adaptable and versatile nature relative to conventional numerical techniques. Unlike standard neural networks reliant solely on data-driven principles, the core effectiveness of PINN resides in its incorporation of physical priors within the network's loss function, encapsulating governing equations alongside initial/boundary conditions, thereby guiding its learning trajectory. By encapsulating underlying physical conservation laws as prior knowledge, PINN offers a synergistic amalgamation of adherence to physical laws and precision in data-driven predictions. Additionally, PINN's versatility extends to scenarios marked by sparse measurement data. Through judicious selection of training samples within convection fields, PINN facilitates the comprehensive reconstruction of all pertinent variables within the flow field, thereby enhancing its applicability in real-world settings beset by data scarcity.

 In the investigation of porous media problems, the intrinsic attributes of PINN render it a universal instrument for confronting the multifaceted challenges prevalent in porous media. Within PINN-based methodologies, the loss function chiefly comprises the $L^2$ norm of the residual associated with the governing equation governing fluid flow within porous media \citep{kashefi2023prediction}. \citet{tartakovsky2018learning} utilized the PINN method to predict the Darcy law governing fluid flux in porous media. \citet{HANNA2022} developed a residual-based adaptive PINN using fully connected networks for two-phase Darcy flow simulation and capturing moving flow fronts. \citet{shokouhi2021physics} incorporated a simple set of two-phase flow equations to impose constraints on two deep learning models used for predicting the simulation response of carbon dioxide storage sites, comparing data-driven and physics-information-driven deep learning models. Despite the PINN model requiring almost twice the training time compared to the data-driven model, it exhibited marginally superior performance in pressure and gas saturation domains. \citet{almajid2022prediction} implemented PINN by an open-source library (DeepXDE)\citep{deepxdelu2021} to solve the forward and inverse two-phase flow problems of the Buckley-Leverett PDE equation, which differs from that of \citet{fraces2020physics} in the construction of the loss function. The documentation and complete code can be found at Github \citep{fraces2020physics}, where a small amount of observed data is included for improved prediction accuracy.
 
 For Darcy's law of fluid flux in porous media, an accurate theoretical model for closing the system of conservation equations is available. However, It is not available for more complex systems, including multiphase flows and transport in porous media \citep{tartakovsky2018learning}. Consequently, to describe unsteady multiphase flows in such situations, it becomes necessary to linearize the equation system. \citet{dieva2024overview} investigated the feasibility of parameterize function application to solve direct and inverse problems of non-Newtonian fluid flows in porous media by using the experimental results of core samples processing in laboratory and field data from a real oil field as examples. 
 
  \begin{figure}[htbp]
  \centering
  \includegraphics[width=\linewidth]{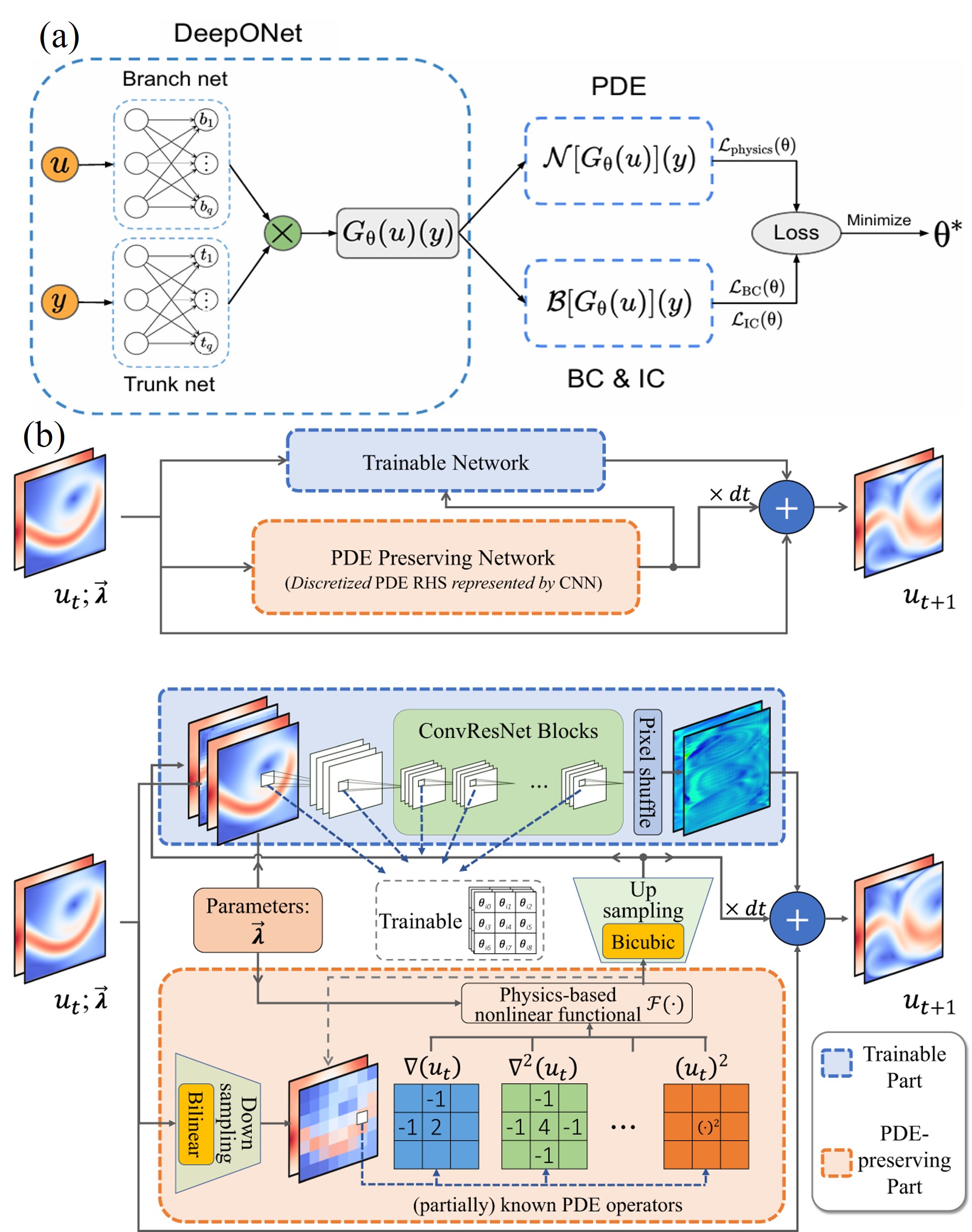}
\caption{Examples of the PINN frameworks of latter versions with various enhancements and extensions. (a) PI-DeepONets \citep{DeepONetswang2021learning}. Reproduced with CC BY 4.0. (b) PPNN \citep{liu2024multiPPNN}. Reproduced with CC BY 4.0.}
  \label{Chapter4.1,Figure.1}
  \end{figure}
 
 Building upon the conventional PINN framework, various researchers have introduced subsequent versions with a plethora of enhancements and extensions. Examples include fPINN \citep{pang2019fpinns}, HFM \citep{HFMraissi2020hidden}, PI-DeepONets \citep{DeepONetswang2021learning}, PPNN \citep{liu2024multiPPNN}, PINN-SR \citep{chen2021PINNSR}, PeRCNN \citep{rao2023PeRCNN}, PINO \citep{PINOazizzadenesheli2024neural}, and some of these frameworks are presented in Fig. \ref{Chapter4.1,Figure.1}.
 Based on the XPINN model proposed by \citet{osti_2282003} as shown in Fig. \ref{Chapter4.1,Figure.2}a, \citet{10.2523} augmented the parallelization ability to solve simple fluid flow scenarios in heterogeneous porous media. Due to the conspicuous discontinuities in hydraulic conductivity, they employed the XPINN concept to partition the complex domain into smaller subdomains, deploying multiple neural networks with optimally selected hyperparameters by enforcing residual continuity conditions on the common interfaces of adjacent subdomains. \citet{zhang2022physics} built a physics-informed convolutional neural network (PICNN) framework grounded on the finite volume method for simulating and predicting single-phase Darcy flows within heterogeneous porous media without labeled data, which can be also extended to simulate two-phase Darcy flows \citep{zhang2023physics} and presented in Fig. \ref{Chapter4.1,Figure.2}b, but only structured grids are allowed. \citet{wang2021efficient,wang2021efficient2,wang2021theory,wang2020deep} proposed a theory-guided convolutional neural network (TgCNN) for predicting subsurface flow problems, forecasting subsurface flow problems, incorporating not only governing equations but also expert knowledge and engineering control into the training process. Subsequent iterations of the proposed TgCNN framework extend to two-phase flow problems in porous media and are employed for efficient inverse modeling.
  \begin{figure}[htbp]
  \centering
  \includegraphics[width=\linewidth]{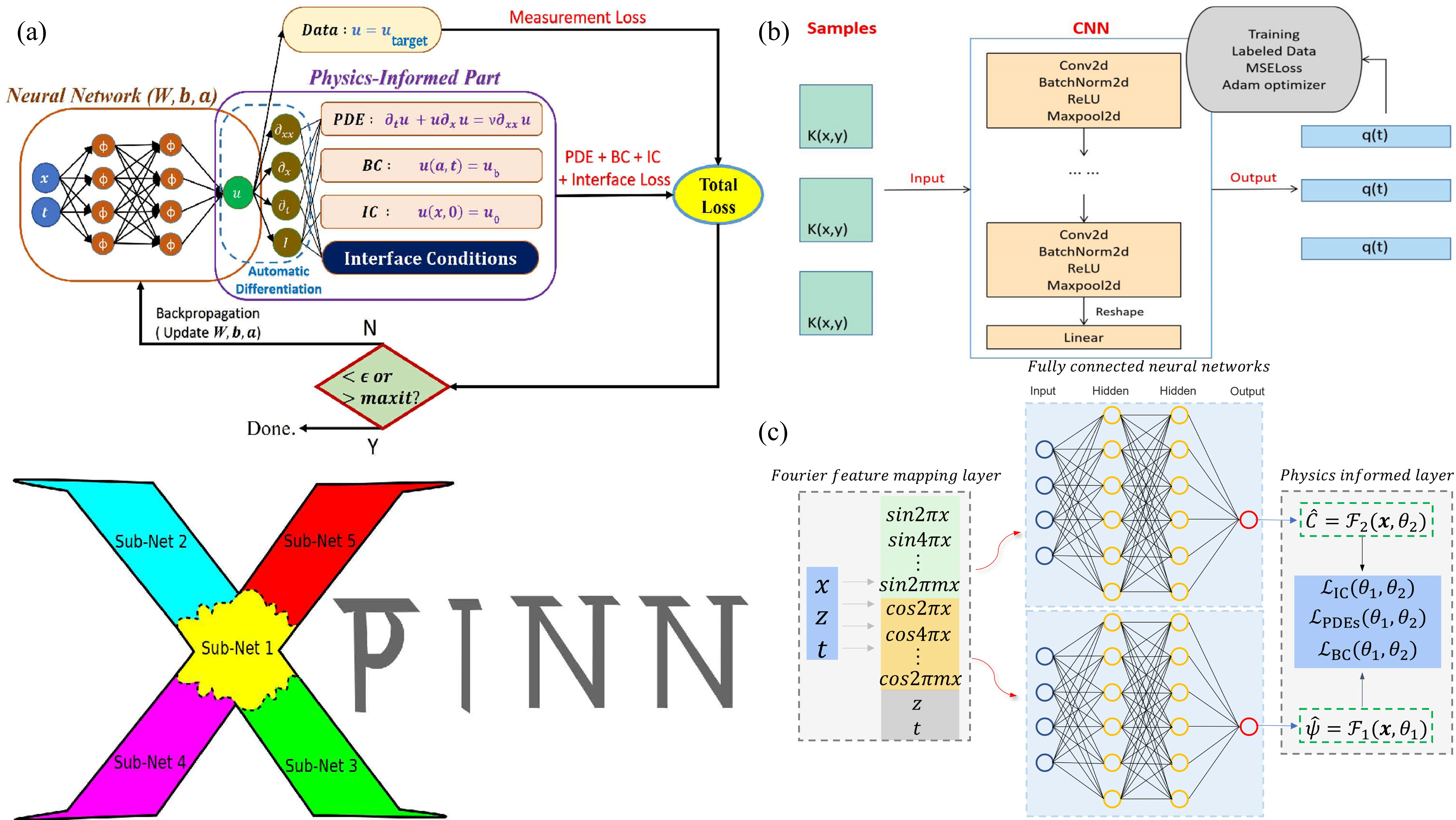}
  \caption{Three extended frameworks of PINN for analyzing flow in porous media. (a) XPINN \citep{osti_2282003}. Reproduced with CC BY 4.0. (b) PICNN \citep{zhang2022physics}. Reproduced with permission from Elsevier. (c) PINN embedded with Fourier basis \citep{du2023fourier}. Reproduced with permission from Elsevier.}
  \label{Chapter4.1,Figure.2}
  \end{figure}
  As shown in Fig. \ref{Chapter4.1,Figure.2}c,\citet{du2023fourier} proposed physics-informed neural networks enriched by Fourier basis embedding, wherein the input space of spatial-temporal coordinates undergoes transformation into spectral space, to examine unsteady carbon dioxide injection processes in porous media. The incorporation of Fourier features notably enhances the prediction accuracy of long-term dynamics and representation of high-frequency fingering patterns to address the spectral bias problem \citep{wang2022bias}, and the proposed approach has enhanced numerical stability even with low resolutions.
  
  Numerous efforts have been devoted to analyze flow in porous media with physics-informed neural networks. Recently, employing PINNs to predict temperature/heat flux fields and effective thermal conductivity of porous media has gained attention \citep{jahanbakhsh2024physics}. 
  \citet{xu2023physics} employed physics-informed neural networks to investigate heat conduction problems in various porous media and the framework as shown in Fig. \ref{Chapter4.1,Figure.3}a. A good performance was shown in predicting temperature fields when compared with numerical calculation results, with an average relative prediction error of 2.49\%, which was presented in Fig. \ref{Chapter4.1,Figure.3}. On the other hand, owing to the absence of data labeling requirements, PINN necessitates considerably less time to develop the proxy model compared to data-driven neural networks, and also achieves a speedup of up to 5 orders of magnitude in prediction over numerical calculations.
  \begin{figure}[htbp]
  \centering
  \includegraphics[width=\linewidth]{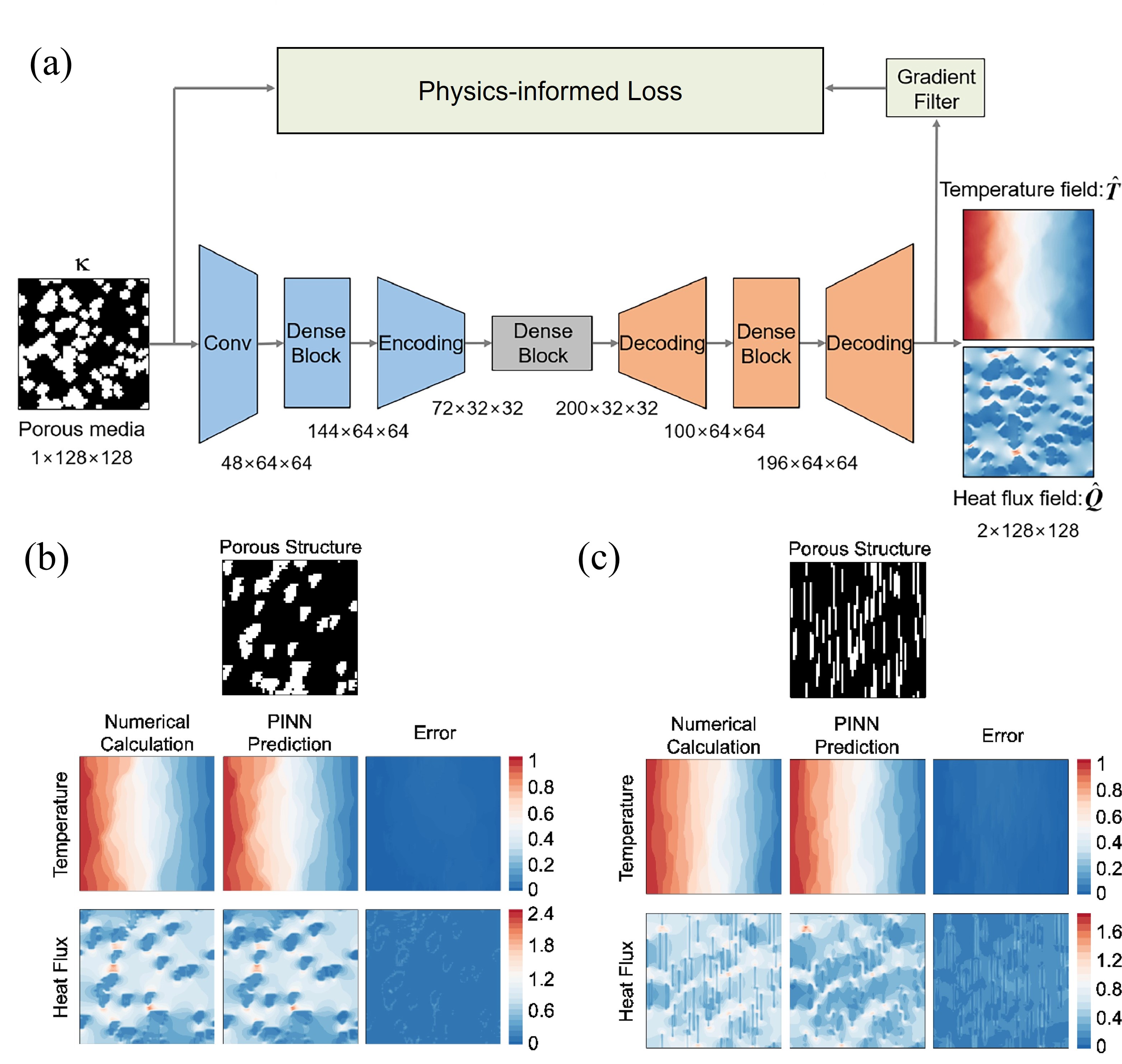}
  \caption{(a) PINN maps the input porous media to desired thermophysical characteristics. The neural network comprises two components. Physics-informed loss functions can be constructed, which constrain the training of the neural network using the physical governing equations without labeled data. (b-c) Temperature/heat flux fields of two porous media samples in the test dataset. The results from numerical calculations, the PINNs, and the errors between them are shown \citep{xu2023physics}. Reproduced with permission from Elsevier.}
  \label{Chapter4.1,Figure.3}
  \end{figure}

\subsection{Hybrid data-driven and physics-informed approaches}
 Physics-Informed Neural Networks (PINNs) have emerged as powerful tools for modeling physical phenomena governed by empirical laws. However, a notable limitation of PINNs is their tendency to incorporate a large number of parameters, often numbering in the millions, to capture the intricacies of these phenomena. These parameters are typically restricted to non-trainable filters within the final layer, leading to potential overfitting issues \citep{kamrava2021}, where the deep learning models excel at fitting training data, but have poor extrapolation capabilities when applied to unseen data \citep{Tetko1995}.
 To address the limitations of deep learning surrogate models, \citet{wang2021physics} presents the PhyFlow-HierCAE model, tailored for predicting structure-dependent pore-fluid velocity fields in rock formations. By embedding the incompressible Navier-Stokes continuity equation and the momentum equation upstream of the Convolutional Autoencoder (CAE), the model gains a fundamental understanding of fluid dynamics principles as presented in Fig. \ref{Chapter4.2,Figure.1}a. Fig. \ref{Chapter4.2,Figure.1}b shows CAE alone cannot predict meaningful flow velocities, both PhyFlow-HierCAE and PhyFlow-CAE successfully yield flow fields similar to the LBM data, while PhyFlow-HierCAE being the most quantitatively accurate. This approach exhibits proficiency in training on limited data and generalizing adeptly to unseen pore structures.
 However, conventional CNN-based approaches for training necessitate labeled data formatted on a Cartesian grid with uniform spacing, potentially leading to unrealistic representations of geometric features near pore-grain boundaries.  
 \begin{figure}[htbp]
  \centering
  \includegraphics[width=\linewidth]{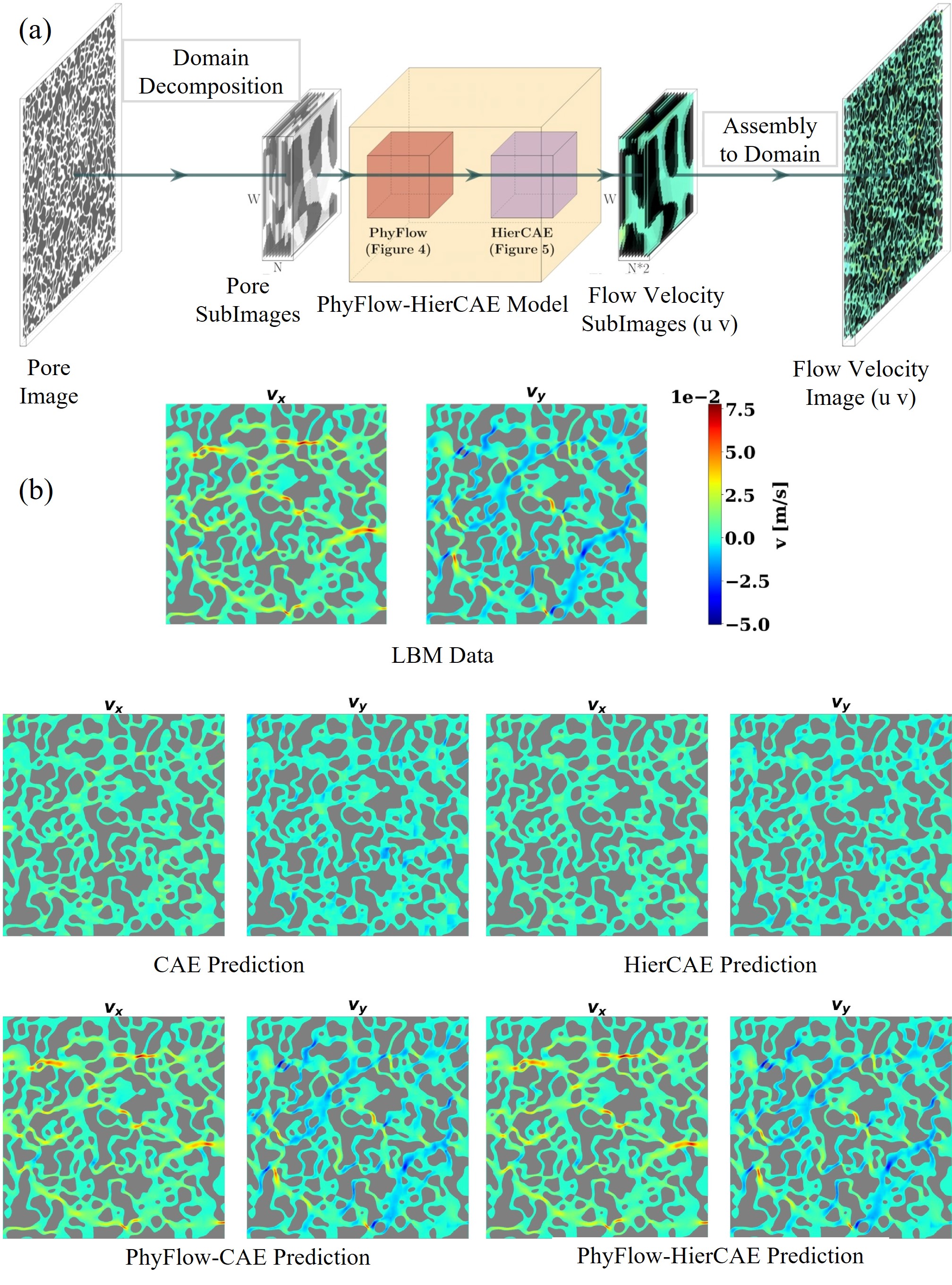}
  \caption{(a) Input/output images and components of the PhyFlow-HierCAE Model. W is the size of the sub-images (windows). N is the number of sub-images. (b) Predictions on the LBM velocity testing data \citep{wang2021physics}. Reproduced with permission from Elsevier.}
  \label{Chapter4.2,Figure.1}
  \end{figure}
 In contrast, \citet{kashefi2022physics} proposed the Physics-Informed PointNet (PIPN) to address these issues, which can allow users to represent the geometry of the pore space and its boundaries smoothly and realistically, and no data interpolation is needed, to predict steady-state Stokes flow of fluids within porous media at pore scales \citep{kashefi2023prediction}. By exclusively utilizing the pore spaces of porous media data as input, PIPN offers adaptable spatial resolution across the physical domain, facilitating to achieve optimal resolution with minimal computational overhead. Moreover, 
 \citet{tartakovsky2020physics} proposed a physics-informed ML method, which used conservation laws in addition to data to train DNNs representing states, space-dependent coefficients, and constitutive relationships. This method enables the prediction of saturated flow in heterogeneous porous media with unknown conductivity and unsaturated flow in homogeneous porous media with an unknown relationship between unsaturated conductivity and capillary pressure. \citet{he2020physics} extended the PINN-based parameter estimation methodology of \citet{tartakovsky2020physics} to predict the transport problem with sparse measurements of hydraulic conductivity, hydraulic head, and solute concentration and refer to this multiphysics-informed neural network approach as MPINN. In which uses the Darcy and advection-dispersion equations together with data to train the deep neural network, the governing equations and the associated boundary conditions are enforced at the “residual” points over the domain. For sparse data, the MPINN approach significantly improves the accuracy of parameter and state estimation as compared to standard DNNs trained with data only. In the subsequent research work, \citet{he2021physics} present the PINN method for solving the coupled advection-dispersion equation (ADE) and Darcy flow equation with space-dependent hydraulic conductivity and test it for one- and two-dimensional forward and backward ADEs for a range of Pe. They demonstrate that assigning larger weights to the residuals of initial and boundary conditions relative to the partial differential equation residuals in the loss function is crucial for obtaining accurate solutions.

\subsection{Machine learning augmented numerical solver}
 The integration of ML models into CFD solvers poses significant challenges due to the disparate nature of their programming languages and optimization domains. While CFD software is typically coded in lower-level languages like Fortran or C++, optimized for CPU computations, ML research often relies on higher-level languages such as Python, tailored for GPU computing. Consequently, for machine learning to gain further widespread application in computational fluid dynamics, there is an urgent need for a new generation of algorithms capable of seamlessly integrating with solver frameworks, operable across CPU, GPU, and TPU platforms. Differentiable solvers are gaining increasing interest in engineering and physical sciences as they promise to bridge the previously mentioned gap between computational physics and machine learning and to ensure the automatic differentiation (AD) of the entire algorithmic representation of the chosen numerical approximation of the PDE evolution law \citep{baydin2018automatic}. Recent advancements in ML have seen neural networks trained using a combination of AD and accompanying equations \cite{freund2019AD,strofer2021AD}, the interest in differentiable simulators has also been driven by the development of powerful general-purpose AD libraries, such as PyTorch \citep{paszke2019pytorch}, TensorFlow \citep{abadi2016tensorflow}, and JAX \citep{bradbury2018jax}, and also numerous specialized libraries, such as JAX-MD \citep{schoenholz2020jaxmd}, JAX-FEM \citep{xue2023jaxFEM}, Phiflow \citep{holl2020phiflow}.
 \citet{bezgin2023jax} developed JAX-Fluids, which can be directly embedded into the framework using existing machine learning models, and uses higher-order numerical methods to compute three-dimensional compressible two-phase flows. JAX-Fluids supports fully differentiable algorithms for end-to-end data-driven model optimization, including both physical and numerical method information, and has recently been expanded to include parallel computation in version 2.0 \citep{bezgin2024jax}. 
 \citet{fraces2020physics} presented an approach relying on a series of deep adversarial neural network architectures with physics-based regularization to simulate two-phase flow in porous problems (Buckley-Leverett), which is highly scalable to leverage CPUs, GPUs and other computing architectures. In subsequent work, they further discussed the limitations of Physics Informed Machine Learning for Nonlinear Two-Phase Transport in Porous Media and further the applicability of this method in forward pure hyperbolic settings \cite{fuks2020limitations,fraces2021physics}.

  \begin{figure}[htbp]
  \centering
  \includegraphics[width=\linewidth]{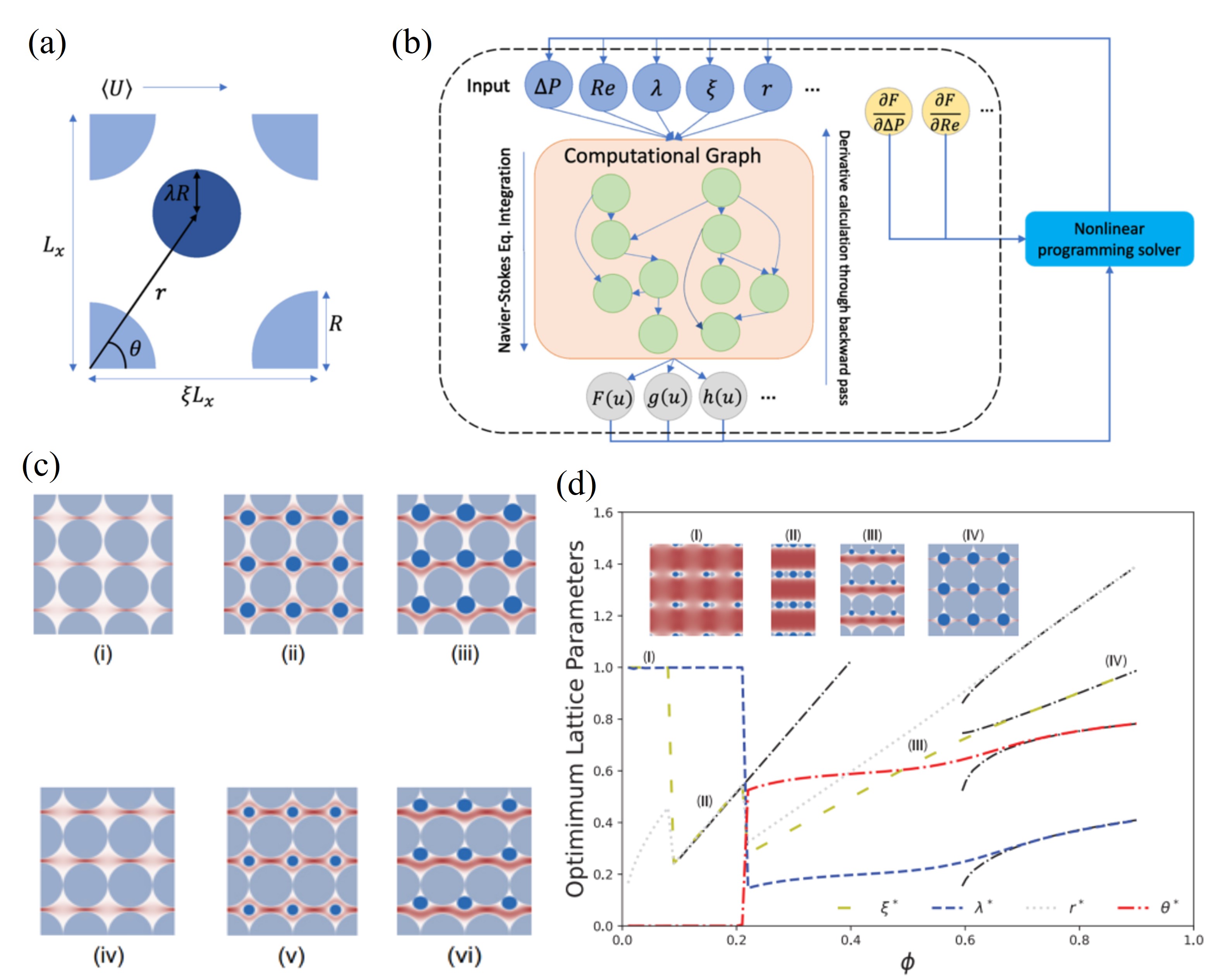}
  \caption{(a) Porous medium configuration. (b) Flowchart of the optimization algorithm using a differentiable Navier-Stokes equation solver. Here F, g, and h are the loss function and the equality and inequality constraints, respectively. (c) Visualization of the different optimization cases: (i) square, (ii) rectangular, (iii) centered square, (iv) off-centered square, (v) centered rectangular, and (vi) off-centered rectangular. (d) Optimized lattice parameters for the off-centered rectangular case \citep{alhashim2024engineering}. Reproduced with permission from American Physical Society.}
  \label{Chapter4.3,Figure.1}
  \end{figure}
  To enhance the accessibility, scalability, parallel computing performance, and multi-physics modeling capabilities of machine learning platforms, \citet{ataei2024xlb} introduce the XLB library, a Python-based differentiable Lattice Boltzmann Method library. This library facilitates the straightforward addition of new boundary conditions and collision models, demonstrating its effective scalability across many GPU devices.
 To maximize flow rate in the direction of applied pressure gradient in porous media of periodically arranged circular rods, \citet{alhashim2024engineering} used JAX-CFD \citep{JAXCFDkochkov2021machine} to solve the two-dimensional periodic velocity field past the array and found that the system with equal particle number density provided an optimum configuration. The periodic structure of porous medium configuration as shown in Fig. \ref{Chapter4.3,Figure.1}a and Fig. \ref{Chapter4.3,Figure.1}b gives a flowchart of the backward differentiation optimization algorithm. In this work, they minimized the impact on the computational cost when increasing the introduction of additional constraints or the number of design parameters. The black dash-dotted curves in Fig. \ref{Chapter4.3,Figure.1}c show a comparison between the numerical and the theory obtained from the optimization of the full flow problem.


\section{Conclusion and perspective}

This review has examined the current advancements in data-driven methods for analyzing flow and transport in porous media. By focusing on three pivotal methodologies—image-based techniques, data-driven flow modeling, and physics-informed machine learning, we have highlighted how these approaches are transforming the field.

Image-based techniques have significantly enhanced the detail and accuracy of images derived from porous structures through innovative methods such as image segmentation, image reconstruction, and super-resolution technologies. These data-driven imaging techniques provide a more precise base for simulation and analysis, allowing for better characterization of flow dynamics within porous media. Recent breakthroughs have integrated these techniques into both traditional and novel modeling frameworks, enhancing the overall fidelity of simulations.
Data-driven flow modeling employs CNNs, U-Nets, GNNs, and other neural network structures to capture the complex patterns of flow and transport dynamics through porous media. These tools have shown substantial promise in offering predictive insights at both microscopic and macroscopic levels, significantly reducing the cost and time associated with traditional modeling methods. By leveraging vast datasets from experiments and simulations, data-driven models can predict critical parameters such as permeability and porosity with moderate to high accuracy.
Physics-informed machine learning is an emergent field that combines the strengths of data-driven methods with fundamental physical principles. Techniques such as PINNs and other hybrid approaches integrate known physical laws directly into the learning process, enhancing the physical plausibility and accuracy of predictions. These methods are particularly effective in addressing complex problems where traditional data is sparse or noisy. Additionally, ML-augmented numerical solvers seamlessly combine simulation and data-driven technologies.

Throughout this review, we have summarized the potential of state-of-art data-driven methods, addressed their limitations, and suggested ways they can be effectively integrated to improve both the fidelity and efficiency of current research. 
Looking forward, several key areas offer significant opportunities for future research and development:

1. Enhanced data integration: Combining data from multiple sources, such as laboratory experiments, field measurements, and high-fidelity simulations, will provide a more comprehensive understanding of porous media systems. Developing robust techniques for data fusion and assimilation will be essential for this integration.

2. Advanced ML algorithms: Continued exploration and development of novel neural network architectures and training techniques will lead to more sophisticated models.

3. Scalable and real-time applications: As these methods mature, developing scalable algorithms and efficient computational frameworks will be crucial for deploying these technologies in practical settings. This will enable real-time monitoring and control, enhancing their applicability in various industrial and environmental contexts.

4. Seamless hybridization of ML and simulations: The seamless integration of ML with CFD through differential fluid dynamics based on automatic differentiation promises to revolutionize the field. Tools like JAX-based, ML-augmented numerical solvers enable this integration by embedding ML models within CFD frameworks. These solvers support fully differentiable algorithms, allowing for end-to-end optimization of multi-physical simulations. This hybrid approach can significantly enhance simulation efficiency and accuracy, unlike the off-line ML in the current stage. 

5. Uncertainty quantification and robustness: Addressing the uncertainties inherent in porous media systems and ensuring the robustness of data-driven models against variations in input data will build trust in these methods. Techniques for uncertainty quantification and model validation will be necessary to enhance the reliability and applicability of data-driven approaches.

In conclusion, the incorporation of image-based techniques, data-driven modeling, and physics-informed machine learning into the study of flow and transport in porous media represents a transformative advancement in the field. By continuing to refine these techniques and explore new applications, researchers can unlock deeper insights and develop innovative solutions to complex thermal fluidic challenges. 

\section*{Acknowledgements}
This study is funded by the National Natural Science Foundation of China (Grant No. 52276013), SFB 1313 from DFG (German Research Foundation, Project No. 327154368).

\bibliographystyle{unsrtnat}
\bibliography{reference}

\end{document}